\documentclass[12pt]{article}
\usepackage{epsfig}
\usepackage{amsmath}
\usepackage{hhline}
\usepackage{amssymb}
\usepackage{times}
\usepackage{cite}

\newlength{\dinwidth}
\newlength{\dinmargin}
\begin{document}
\newcommand{\xpomlo}{3\times10^{-4}}  
\newcommand{\xpomup}{0.05}  
\newcommand{\modP}{|{\bf P}|}
\newcommand{\mode}{|{\bf e}|}
\newcommand{\pom}{{I\!\!P}}
\newcommand{\fiidiii}{F_2^{D(3)}}
\newcommand{\fiidiiiarg}{\fiidiii\,(\beta,\,Q^2,\,x)}
\newcommand{\n}{1.19\pm 0.06 (stat.) \pm0.07 (syst.)}
\newcommand{\nz}{1.30\pm 0.08 (stat.)^{+0.08}_{-0.14} (syst.)}
\newcommand{\fiidiiiful}{F_2^{D(4)}\,(\beta,\,Q^2,\,x,\,t)}
\newcommand{\fiipom}{\tilde F_2^D}
\newcommand{\ALPHA}{1.10\pm0.03 (stat.) \pm0.04 (syst.)}
\newcommand{\ALPHAZ}{1.15\pm0.04 (stat.)^{+0.04}_{-0.07} (syst.)}
\newcommand{\fiipomarg}{\fiipom\,(\beta,\,Q^2)}
\newcommand{\pomflux}{f_{\pom / p}}
\newcommand{\nxpom}{1.19\pm 0.06 (stat.) \pm0.07 (syst.)}
\newcommand {\gapprox}
   {\raisebox{-0.7ex}{$\stackrel {\textstyle>}{\sim}$}}
\newcommand {\lapprox}
   {\raisebox{-0.7ex}{$\stackrel {\textstyle<}{\sim}$}}
\newcommand{\pomfluxarg}{f_{\pom / p}\,(x_\pom)}
\newcommand{\dsf}{\mbox{$F_2^{D(3)}$}}
\newcommand{\dsfva}{\mbox{$F_2^{D(3)}(\beta,Q^2,x_{I\!\!P})$}}
\newcommand{\dsfvb}{\mbox{$F_2^{D(3)}(\beta,Q^2,x)$}}
\newcommand{\dsfpom}{$F_2^{I\!\!P}$}
\newcommand{\gap}{\stackrel{>}{\sim}}
\newcommand{\lap}{\stackrel{<}{\sim}}
\newcommand{\fem}{$F_2^{em}$}
\newcommand{\tsnmp}{$\tilde{\sigma}_{NC}(e^{\mp})$}
\newcommand{\tsnm}{$\tilde{\sigma}_{NC}(e^-)$}
\newcommand{\tsnp}{$\tilde{\sigma}_{NC}(e^+)$}
\newcommand{\st}{$\star$}
\newcommand{\sst}{$\star \star$}
\newcommand{\ssst}{$\star \star \star$}
\newcommand{\sssst}{$\star \star \star \star$}
\newcommand{\tw}{\theta_W}
\newcommand{\sw}{\sin{\theta_W}}
\newcommand{\cw}{\cos{\theta_W}}
\newcommand{\sww}{\sin^2{\theta_W}}
\newcommand{\cww}{\cos^2{\theta_W}}
\newcommand{\trm}{m_{\perp}}
\newcommand{\trp}{p_{\perp}}
\newcommand{\trmm}{m_{\perp}^2}
\newcommand{\trpp}{p_{\perp}^2}
\newcommand{\alp}{\alpha_s}

\newcommand{\alps}{\alpha_s}
\newcommand{\sqrts}{$\sqrt{s}$}
\newcommand{\LO}{$O(\alpha_s^0)$}
\newcommand{\Oa}{$O(\alpha_s)$}
\newcommand{\Oaa}{$O(\alpha_s^2)$}
\newcommand{\PT}{p_{\perp}}
\newcommand{\JPSI}{J/\psi}
\newcommand{\GP}{\gamma p  \rightarrow }
\newcommand{\Gg}{\gamma g  \rightarrow }
\newcommand{\sh}{\hat{s}}
\newcommand{\uh}{\hat{u}}
\newcommand{\MP}{m_{J/\psi}}
\newcommand{\PO}{I\!\!P}
\newcommand{\Pvec}{\mbox{\boldmath $P$}}
\newcommand{\pvec}{\mbox{\boldmath $p$}}
\newcommand{\qvec}{\mbox{\boldmath $q$}}
\newcommand{\Vvec}{\mbox{\boldmath $V$}}
\newcommand{\Pvecprime}{\mbox{\boldmath $P'$}}
\newcommand{\pvecprime}{\mbox{\boldmath $p'$}}
\newcommand{\pvecfoot}{\begin{footnotesize}{\mbox{\boldmath $p$}}\end{footnotesize}}
\newcommand{\xbj}{x}
\newcommand{\xpom}{x_{\PO}}
\newcommand{\ttbs}{\char'134}

\begin{titlepage}
\begin{flushleft}
{\tt DESY 00-137}\hfill {\tt ISSN 0418-9833} \\
{\tt \today \\}
\end{flushleft}
\vspace*{2.5cm}
\begin{center}
\begin{huge}
{\bf Photon Structure \\ and \\
Quantum Fluctuation}\footnote{Invited presentation at the discussion meeting ``The Quark Structure of Matter'', Royal Society, London, May 24th and 25th 2000; to appear in Philosophical Transactions of the Royal Society of London (Series A: Mathematical, Physical and Engineering Sciences).}  \\
\end{huge}
\vspace{2cm}
\begin{Large}
John Dainton  \\
\end{Large}
\vspace{0.6cm}
Oliver Lodge Laboratory, \\
Department of Physics, \\
The University of Liverpool\footnote{Present address: Department of Physics, University of Liverpool, Oxford Street, Liverpool L69 7ZE; e--mail: dainton@mail.desy.de} \\
\vspace{0.5cm}
\end{center}
\vspace{2cm}
\begin{abstract}
\noindent
Photon structure derives from quantum fluctuation in quantum field theory to fermion and anti--fermion, and has been an experimentally established feature of electrodynamics since the discovery of the positron. In hadronic physics, the observation of factorisable photon structure is similarly a fundamental test of the quantum field theory Quantum Chromodynamics (QCD). An overview of measurements of hadronic photon structure in $e^+e^-$ and $ep$ interactions is presented, and comparison made with theoretical expectation, drawing on the essential features of photon fluctuation into quark and anti--quark in QCD.
  \end{abstract}
\end{titlepage}

\section{Introduction}
\subsection{Historical Context}
The structure of the proton, the simplest atomic nucleus and thereby the simplest piece of matter that we have readily available, has for decades been central to the development of our understanding of the universe. Following the experimental principles first established with the birth of nuclear physics and Rutherford's classic experiment with $\alpha -$particles~\cite{Rutherford}, an understanding now exists of the structure of matter in terms of the fundamental building blocks, quarks ($q$) and gluons ($g$), or collectively partons. In particular a proton has structure built from three (valence) constituents of different flavour, 2 up--quarks ($u$) and one down--quark ($d$), and their associated flavour singlet sea~\cite{Taylor}. This structure is maintained because the fermions, the quarks, have ``colour'' and interact with each other in a manner described by a non--abelian vector field theory, Quantum Chromodynamics (QCD)~\cite{QCD}, in which the spin--1 field quanta are the gluons. The gluons also have colour. Unlike atoms and molecules, the masses of hadrons are much greater than the masses of all of their constituents, so that the dynamic equilibrium of valence quarks and sea quarks making up this structure is necessarily relativistic, and any ``snapshot'' of proton structure also reveals a substantial density of field quanta, gluons~\cite{Foster}.

Structure is probed by means of scattering experiments in which the distance scale which is resolved is specified by the energy and momentum transfers in the interaction. In terms of a Lorentz $4-$momentum $q$ describing these energy and momentum transfers, the characteristic temporal and spatial resolutions follow from the Heisenberg Uncertainty Principle, and are $1/q_0$ and $1/|\qvec|$ respectively\footnote[1]{\label{foot:Heisenberg}In this paper rational units in which $\hbar=c=1$ are used except where doing so obscures understanding; thus here the Heisenberg Uncertainty Relations read $\Delta E\Delta T\sim 1$ and $\Delta p\Delta x\sim 1$. Also the $4-$momentum metric $q=(q_0,\qvec)$ and $q^2=q_0^2-\qvec^2$ is taken. As a numerical rule of thumb $\Delta p ({\rm MeV})\Delta x ({\rm fm})\sim 200$ MeV fm.}. This principle, which is manifest in the Fourier Transform relationship between the functions describing the structure being probed and the scattering angular distribution, underpins the ethos of most of High Energy Physics. New levels of structure in matter and interactions have characterised each major step forward. At each step it has been possible to establish an understanding in terms of a Quantum Field Theory (QFT), such as QCD, or to unify previous observations, such as Quantum Electrodynamics (QED) and weak ($\beta-$decay) interactions, into a single QFT, Electroweak Theory with massive gauge bosons $W$ and $Z$ alongside the massless photon ($\gamma$), leptons $e$, $\mu$ and $\tau$, and quarks of different flavour. The role therefore of QFT in our understanding of the structure of matter is seminal, and interpretation of results in hadron physics and hadronic structure in terms of QCD is nowadays automatic.

\subsection{Structure due to Quantum Fluctuation}
\label{sub:pointlikestructure}
\begin{figure}[hbt]
  \begin{center}
\epsfig{file=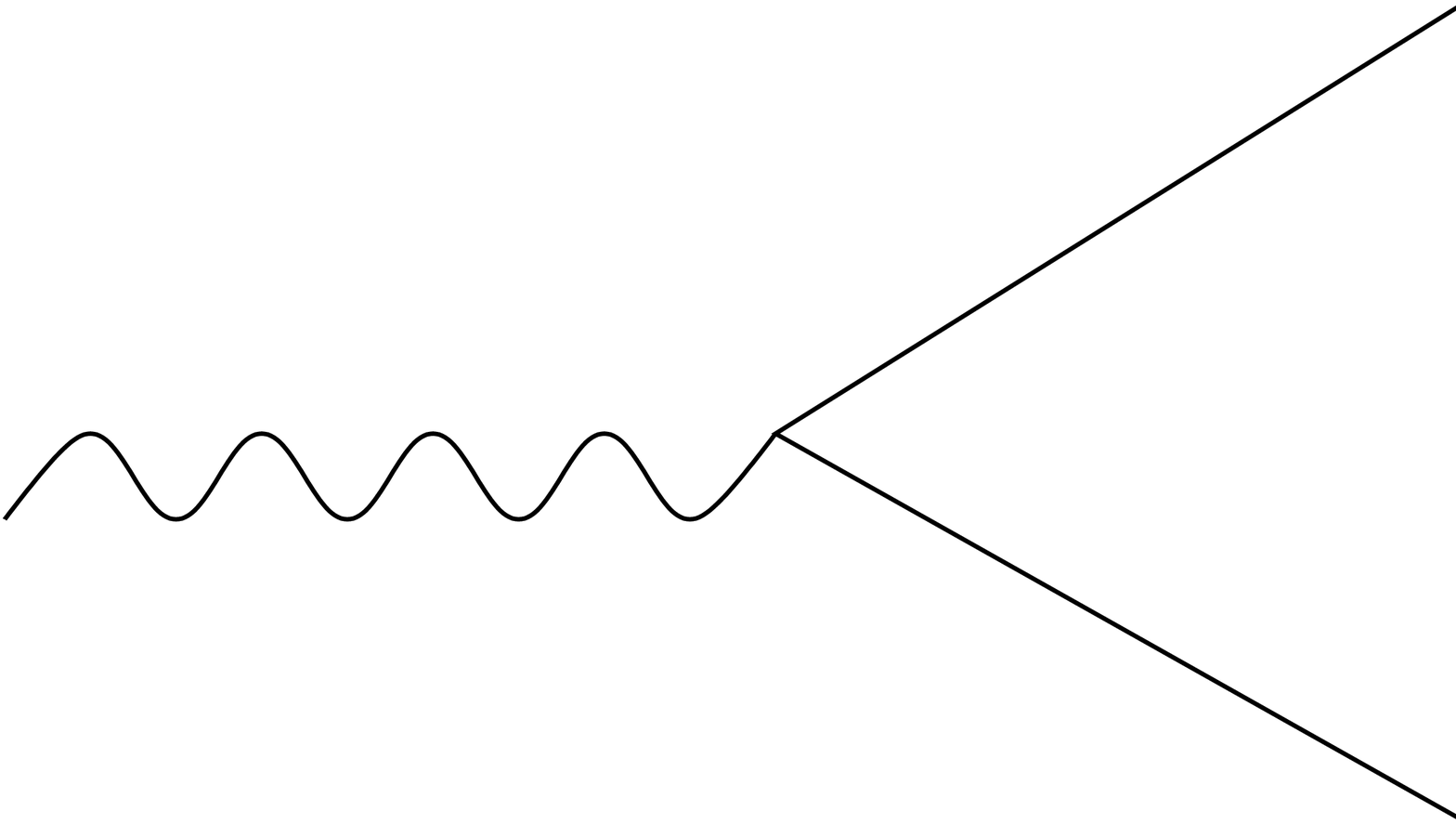,width=0.24\textwidth}
\hfill
    \epsfig{file=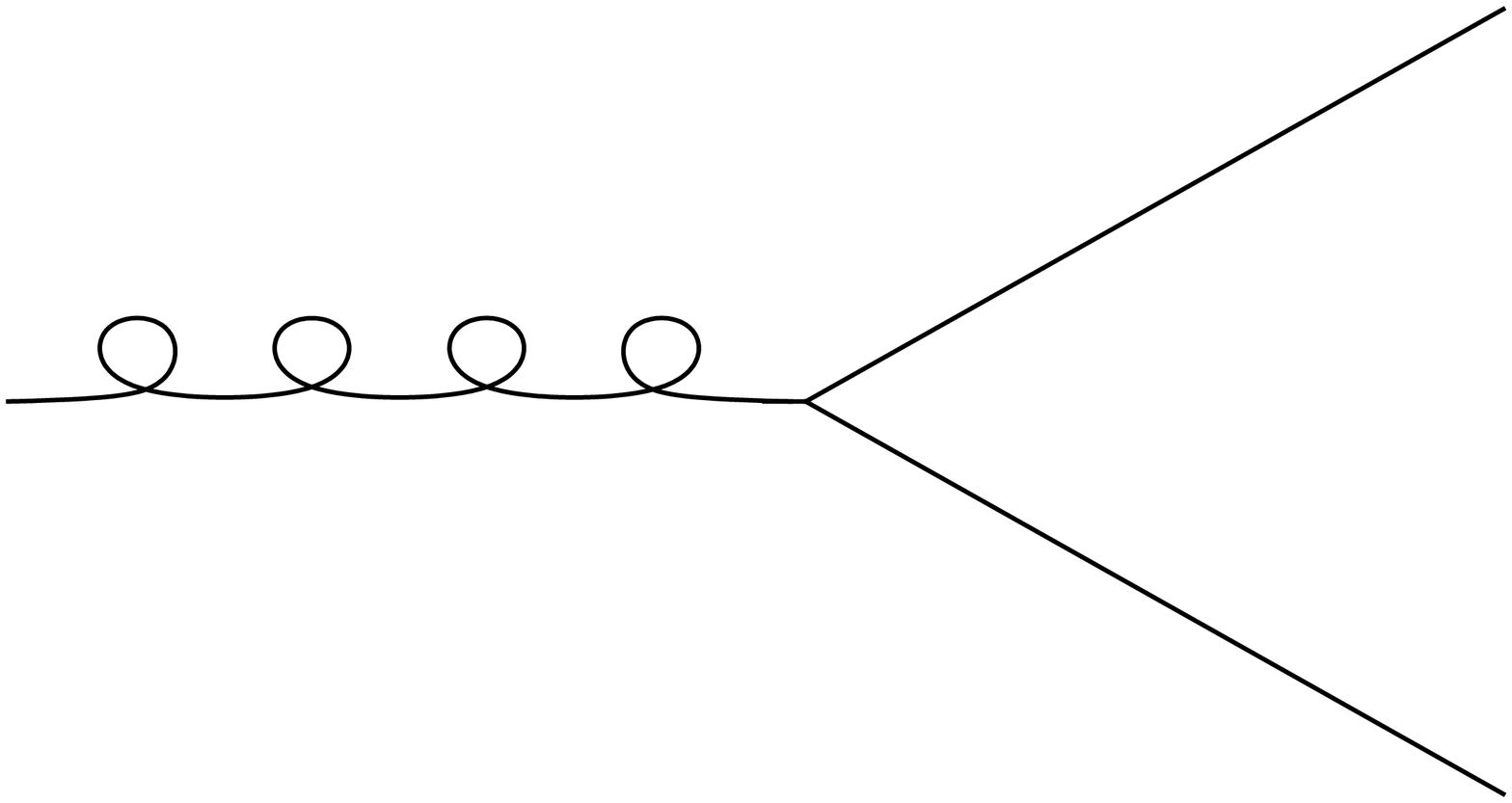,width=0.24\textwidth}
\hfill
    \epsfig{file=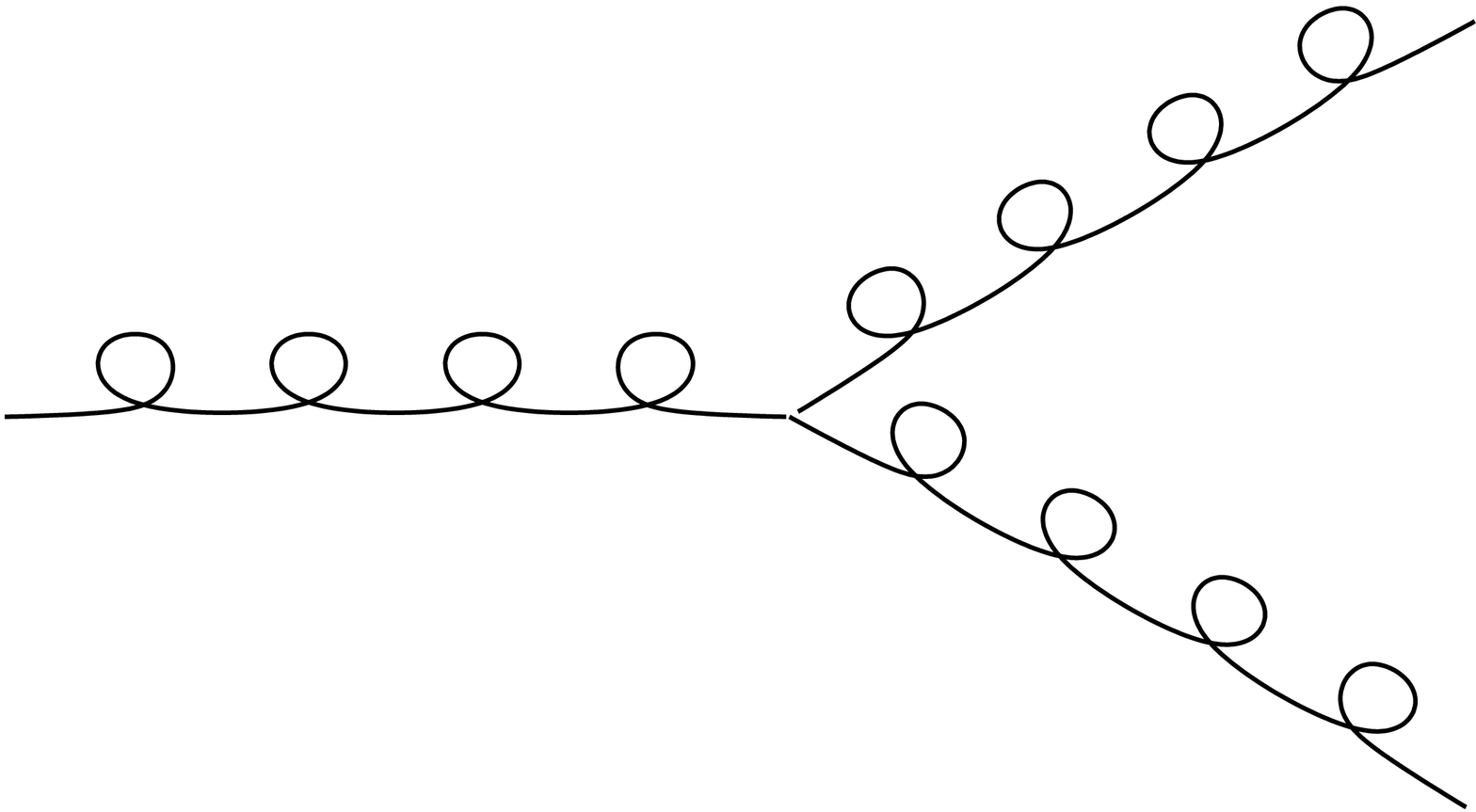,width=0.24\textwidth}
\hfill
    \epsfig{file=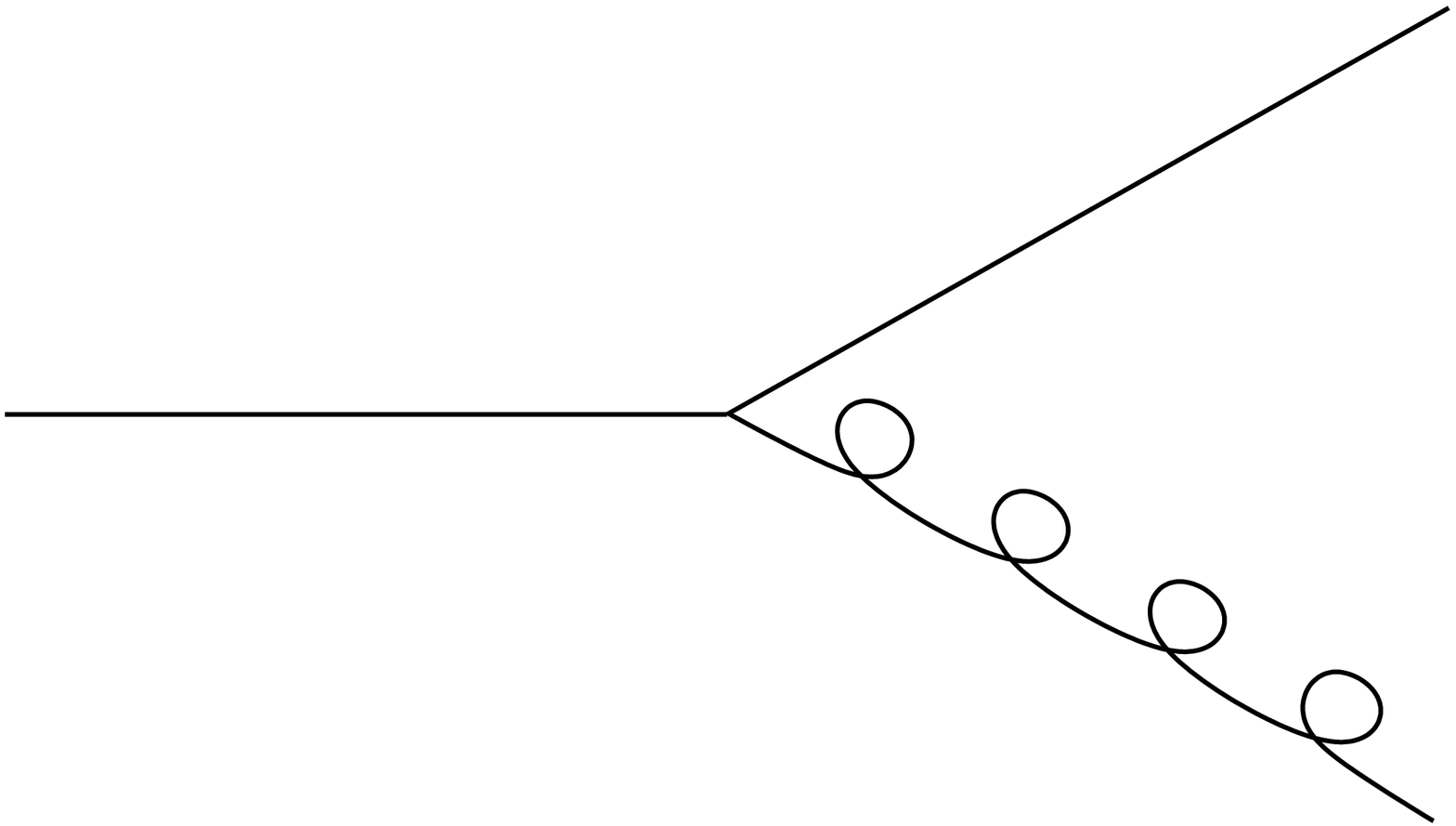,width=0.24\textwidth}
\caption{\footnotesize Schematic diagrams illustrating the simplest fluctuation, or splitting, of quanta in Quantum Field Theory; in Quantum Electrodynamics (QED) a photon (real or virtual) can split into electron and positron, or muon and anti--muon, or tau and anti--tau; in Quantum Flavour Dynamics (QFD) the pair could also be $W^+$ and $W^--$bosons or quark and anti--quark; the variable $t$ specifies the space--like invariant mass squared of the off--shell splitting product and the variable $y$ specifies the ``inelasticity'' (see text) of the splitting; fluctuations in Quantum Chromodynamics (QCD): b) quark--antiquark, and c) gluon--gluon, pair production by a single gluon, and d) quark ``Bremsstrahlung'' to quark and gluon.}
    \label{fig:bubble}
\end{center}
\vspace{-6.2cm}
\hspace{0.2cm}
a)\hspace{1.3cm}\begin{footnotesize}on shell\end{footnotesize}
\hspace{1.1cm}
b)
\hspace{3.7cm}
c)
\hspace{3.7cm}
d)
\vspace{0.1cm}

\begin{footnotesize}photon \hspace{2cm}$p'$\hspace{1cm}gluon\hspace{3.5cm}gluon\hspace{3.5cm}quark
\vspace{0.5cm}

$p$

\vspace{-0.3cm}\hspace{1.1cm}off shell $p_e$

\vspace{-0.1cm}\hspace{1.7cm}space-like
\vspace{4.6cm}\end{footnotesize}
\end{figure}

In any QFT, the existence of interactions means also that the quanta themselves have structure. This structure derives from the fact that the mutual coupling between field quanta will make possible fluctuation, or splitting, of any quantum over a limited time/distance into two (or more in higher order) quanta (figure~\ref{fig:bubble}). For example in QED a photon can fluctuate for a short time into an electron--positron ($e^+e^-$) pair or an electron can fluctuate into an electron and a photon ($\gamma$).

In Lorentz covariant terms, one or more of the quanta into which fluctuation occurs are ``virtual'' and therefore ``off--mass--shell'', or just ``off--shell''. In QCD/QED, the probability of a fluctuation is proportional to $1/(m_e^2-t)$ where $t=p_e^2=(p-p')^2$ is the space--like mass squared of the off--shell quantum and $m_e$ is its on--shell mass (figure~\ref{fig:bubble}a). Both time orderings, the emission of a quantum and absorption of a quantum, are of course included in this covariant probability. However in a Lorentz frame where the $3-$momentum of the parent becomes large, the expression $1/(m_e^2-t)$ is dominated by the term arising from emission. This high momentum limit is often described as the ``light--cone limit'' because the space--like $4-$momenta tend towards the light cone as masses and transverse momenta become small relative to the parent $3-$momentum, or more loosely it is referred to as the picture in the ``infinite momentum frame''~\cite{Feynman}. 
 
One of the consequences of this ``fluctuation--driven structure'' is that the definition of a quantum in an interaction is arbitrary because the view one has of it depends on the spatial precision, that is momentum transfer scale, available. Thus probing fluctuation--driven structure by means of a scattering experiment will continue to reveal new detail with increasing spatial resolution, until there is a breakdown of the theory and a discovery! Furthermore, the consequences of unresolved quantum structure because of inadequate spatial precision are that the coupling and mass of the quantum are observed to vary with the momentum transfer scale of the interaction. This ``renormalisation'', as it is called, was first observed for QED in precision measurements by Lamb and Retherford of the r.f. spectroscopy of atomic hydrogen~\cite{LambRetherford}. Nowadays renormalisation forms the basis of probing the Standard Model in $e^+e^-$ physics at LEP at CERN, Geneva, and is manifest most dramatically in the form of ``running coupling constants'' in QFT. The most well known of these is the QCD, or strong, coupling constant $\alpha_S(q^2)$ which decreases with increasing momentum transfer squared scale $Q^2=-q^2$. Thereby QCD is an ``asymptotically free'' theory in which the short distance interactions of quarks and gluons are weak ($\alpha_S$ small when $Q^2=-q^2$ large) but the long distance interactions are strong ($\alpha_S$ large when $Q^2$ small)~\cite{QCD}.

\subsection{Kinematics of Structure}
\label{sec:kin}
The kinematics of structure in terms of quantum splitting of a parent particle into two product particles, be it in terms of a QFT or arising in the specification of the structure of the parent, is relatively straightforward. Two (Lorentz covariant) variables\footnote{Here the possibility of spin polarisation and the need to specify azimuthal angle is explicitely ignored.} are chosen. To at least one of the products is assigned an off--shell, space--like invariant mass squared 
\begin{equation}
t=(p-p')^2
           \label{eq:tdef}
\end{equation} 
so that energy and momentum are conserved even when the other product particle and the parent particle are massless\footnote{In principle {\em one or both} of the products of the splitting can be off mass shell, but the probability for the latter to happen is low and only rarely of importance.}. Here $p$ is the parent $4-$momentum and $p'$ is the on-shell product $4-$momentum (figure~\ref{fig:bubble}), for purposes here taken to be timelike and massless, $(p')^2=0$ and $p'_0=|\pvecprime|$. The second variable is chosen to be the Lorentz invariant combination 
\begin{equation}
y=\frac{(p-p').V}{p.V}=1-\frac{p'.V}{p.V}
           \label{eq:y}
\end{equation} 
where $V$ is any $4-$vector used to specify the Lorentz frame in which $y$ is to be evaluated. Given that the splitting is observed as part of an interaction of the parent particle with another particle, $V$ is taken naturally to be the $4-$vector of the other particle in the chosen frame.

Experimentally the two variables which naturally specify the kinematics are transverse momentum $p_T$ and longitudinal momentum $p'_z$ of the on--shell splitting product, defined with respect to the parent's momentum vector as the $z-$axis. In a high energy interaction it is advantageous to work with rapidity\footnote{To avoid confusion caused by the danger of a plethora of variables termed $y$, rapidity is here written $\eta$ and not its more traditional $y$. For a particle of energy $E$, mass $m$ and $z-$component of momentum $P_z$ rapidity $\eta=\frac{1}{2}\ln\frac{E+P_z}{E-P_z}\equiv\ln\frac{E+P_z}{\sqrt{m^2+P_T^2}}$. Note also that often $\eta=-\frac{1}{2}\ln\tan\frac{\theta}{2}$ is taken to denote pseudo--rapidity where $\theta=\arctan\frac{P_T}{P_z}$ is the polar angle of the particle. Rapidity tends to pseudo--rapidity as the particle becomes relativistic.} $\eta$, rather than $p'_z$. Physical insight into $t$ and $y$, in terms of their relationship to the measureds $p_T$ and $\eta$, depends on the Lorentz frame of reference. Because one here considers the structure of particles, photons, which can be on the light cone, the appropriate frame is that in which the parent has very large $3-$momentum $\pvec$ (Feynman's ``infinite momentum frame'' for the parent) such that $p_z$ defines the $z-$axis and one can work to leading order in $p^2/p_0^2$. Then
\begin{equation}
\Delta \eta =\Delta\ln(1-y).
           \label{eq:ytorapidity}
\end{equation}
and a change of $y$ specifies a change of the rapidity of the on--shell splitting product. Also
\begin{equation}
t=yp^2-\frac{p_T^2}{1-y}
           \label{eq:ttoPTlightcone}
\end{equation}
in which one can see the extent to which the $4-$momentum squared $t$ is manifest, or not, in the transverse $3-$momentum squared $p_T^2$ of the splitting. 
In this frame equation~\ref{eq:y} can now be written
\begin{equation}
y\approx 1-\frac{p_0'}{p_0},
           \label{eq:yinfmomentum}
\end{equation}
and, to an accuracy which ignores terms ${\cal O}(\frac{p_T^2}{p_0p'_0})+{\cal O}(\frac{p_T^2}{p_0^2})$, $y$ and $1-y$ are now seen to specify the fraction of the parent momentum (or equivalently in this limit parent energy) carried by the products.

Thus the kinematics of quantum splitting are specified with the variables transverse momentum $p_T$ and rapidity $\eta$ of the on--shell quantum. These variables are related to the virtuality $t$ and inelasticity $y$ respectively of the off--shell quantum. Thereby they specify the transverse and longitudinal characteristics of the splitting.

\subsection{The Spatial Extent of fluctuation--driven Structure}
\label{sec:dimension}

The spatial extent of quantum fluctuation depends on the energy/momentum imbalance observed in a particular Lorentz frame and what can be tolerated by the Heisenberg Uncertainty Principle. The most intuitive insight is again obtained by continuing in the high momentum limit of the parent quantum.

The ``lifetime'' $\tau$ of a fluctuation is specified by the energy imbalance $\Delta E$ when all particles are on--shell. Using equation~\ref{eq:yinfmomentum} and the notation in figure~\ref{fig:bubble}, straightforward evaluation gives 
$$\Delta E=p_{0}(0)-p_{0}-[p_{e0}(m_e^2)-p_{e0}]\approx\frac{m_e^2+yp^2-t}{2yp_{e0}}.$$ 
Here $p_{0}(0)$ and $p_{e0}(m_e^2)$ are the energies of the off--shell quanta with $3-$momenta $\pvec$ and $\pvec_e$ evaluated with on--shell mass squareds $0$ (rather than $p^2$) and $m_e^2$ (rather than $t$) respectively. As long as $2yp_{e0}\gg \sqrt{m_e^2+yp^2-t}$, $\Delta E$ is small and the temporal dimension of the fluctuation ($\tau\sim 1/\Delta E$) is dilated and therefore large. The probe, whose time--base ($1/q_0$) is easily fixed to be much shorter than $\tau$, thus takes a ``snapshot'' of the structure, and the fluctuation lives for much longer than any typical interaction time. Therefore it becomes possible to think in terms of the interaction of the fluctuation rather than of the parent quantum itself. Given the near light--cone velocity of the parent, it is useful to quantify this in terms of a length ($\beta=|\pvec_e|/p_{e0}\sim 1$) $$\Delta z=\frac{2yp_{e0}}{m_e^2-t-yp^2}$$ of the quantum fluctuation. This is known as the Ioffe length (or time)~\cite{Ioffe}; in the literature it is quoted for an on--shell parent ($p^2=0$).

The Ioffe length should not be confused with the longitudinal dimension of the fluctuation. The parton model picture of a hadron in the infinite momentum frame has nearly all its size manifest in the dimensions perpendicular to the direction of its high momentum. Fluctuation--driven structure is similarly Lorentz contracted.

The transverse dimension arises because of the $p_T$ in the fluctuation and the degree to which the splitting quanta separate as the Ioffe time elapses. Writing equation~\ref{eq:ttoPTlightcone} in terms of the angle of the on--shell fluctuation product $\theta$ using $p_T^2=(p_0')^2\theta^2=(1-y)^2p_0^2\theta^2$ in the small angle approximation appropriate for a high momentum frame, then
$$\theta=\sqrt{\frac{yp^2-t}{1-y}}\cdot \frac{1}{p_0}.$$
Immediately the transverse dimension is estimated to be
\begin{equation}
\Delta x\approx \frac{2yp_{e0}}{m_e^2+yp^2-t}\cdot \theta=\frac{2y}{\sqrt{1-y}}\cdot \frac{\sqrt{yp^2-t}}{m_e^2+yp^2-t}=\frac{2y}{\sqrt{1-y}}\cdot \frac{\sqrt{-t'}}{m_e^2-t'}
\label{eq:transversesize}
\end{equation}
where the variable $t'=t-yp^2$ is introduced as the extent to which $|t|$ exceeds its minimum value $|t_{\rm min}|=|yp^2|$ (equation~\ref{eq:ttoPTlightcone}). When $|t'|$ is much larger than both the on--shell mass squared $m_e^2$ and the virtuality of the parent $|yp^2|$, the transverse dimension is $$\Delta x\approx\frac{2y}{\sqrt{1-y}}\frac{1}{\sqrt{-t'}}.$$ When $|t'|$ is small relative to $m_e^2$ or $|yp^2|$, 
\begin{equation}
\Delta x\approx\frac{2y}{\sqrt{1-y}}\frac{1}{m_e}\,\,\,\,\,(m_e^2\gg |yp^2|)\,\,\,\,\,{\rm or}\,\,\,\,\,\Delta x\approx 2\sqrt{\frac{y}{1-y}}\frac{1}{P}\,\,\,\,\,(m_e^2\ll |yp^2|)
\label{eq:size}
\end{equation}
and the transverse dimension is set by the larger of the two. 

\subsection{Manifestation of Structure in Interactions}
\label{sec:manifestation}

\subsubsection{Formalism}
\begin{figure}[hbt]
  \begin{center}
\hspace*{0.4cm}\epsfig{file=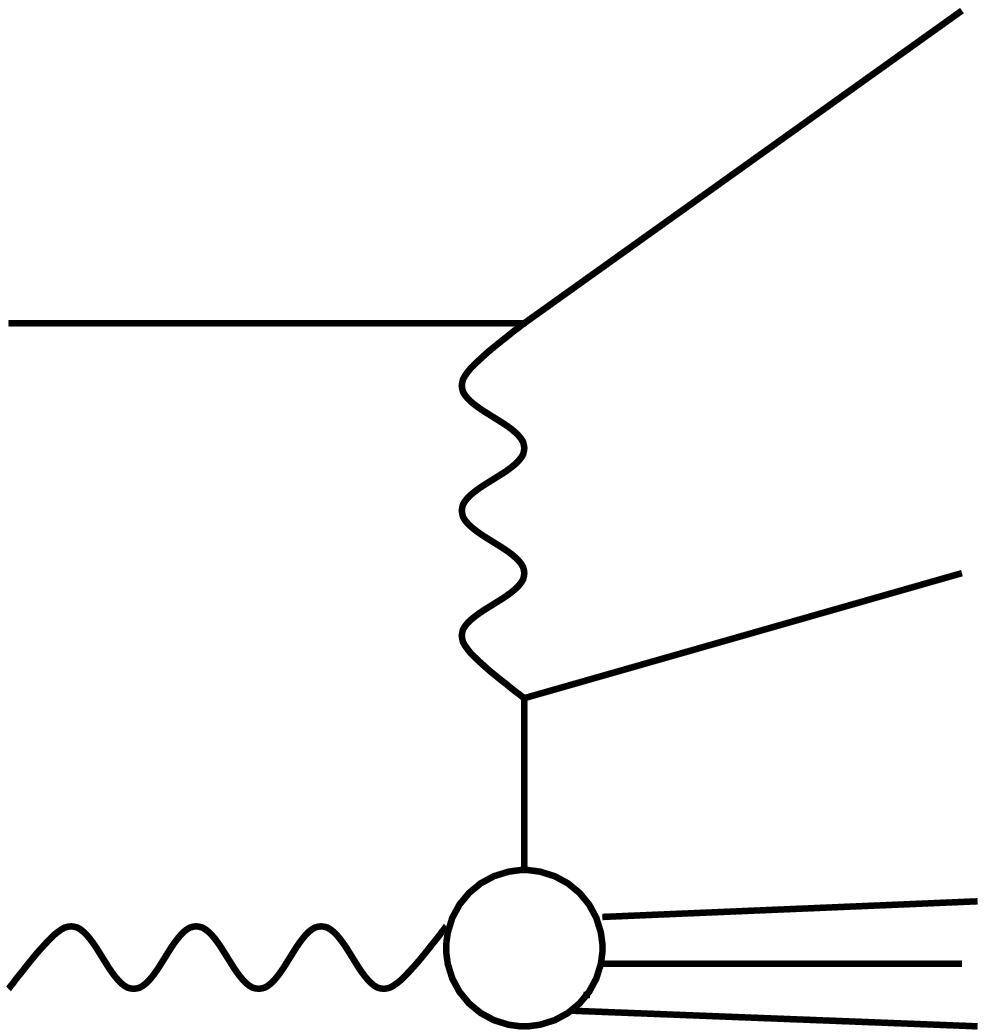,width=0.42\textwidth}\hfill\epsfig{file=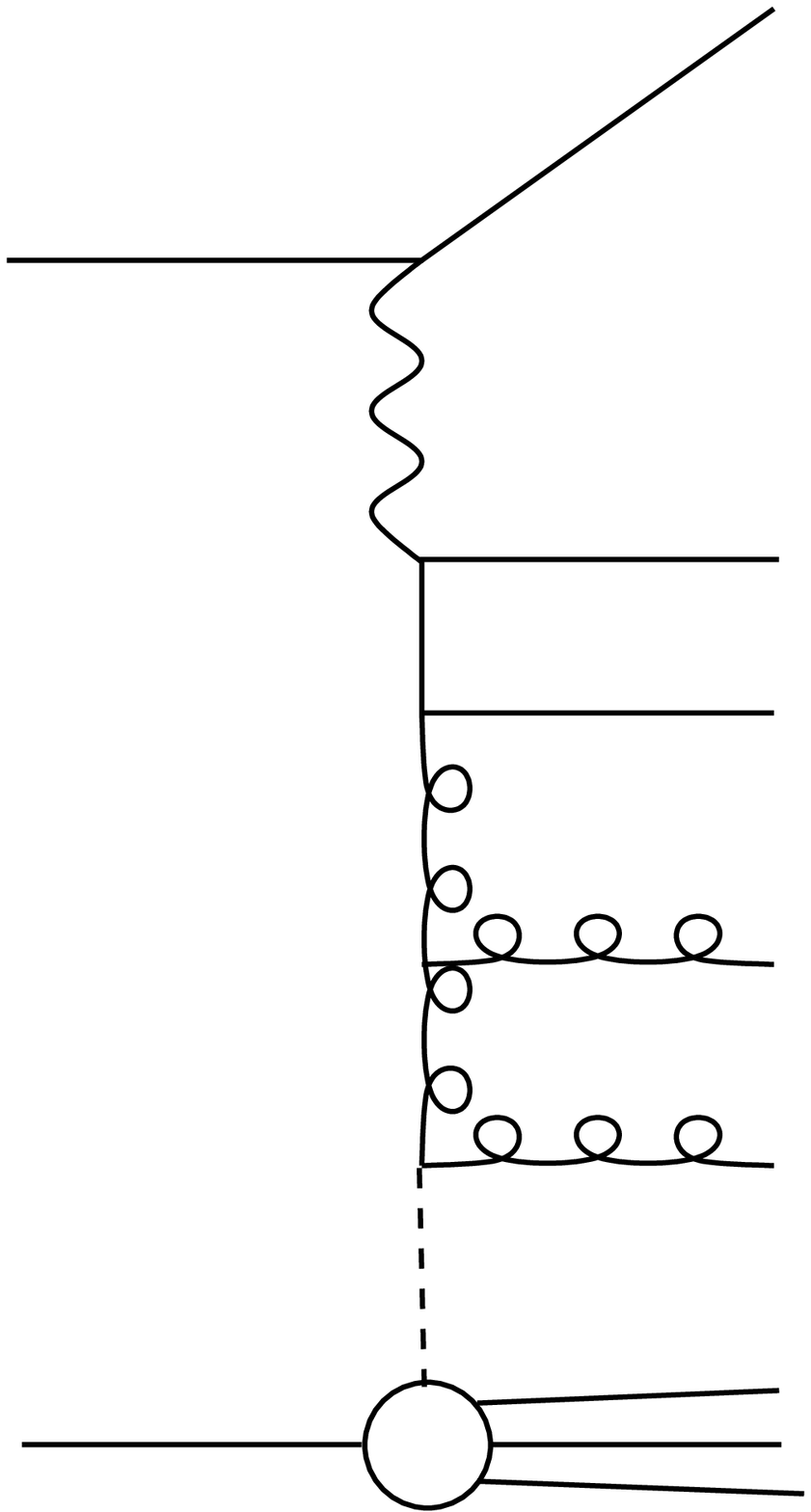,width=0.24\textwidth}\hspace*{2cm}
\caption{\footnotesize a) Schematic diagram illustrating how in an interaction the structure of a particle (here a photon $\gamma$) is probed with a ``point--like'' particle (electron $e$); the scale of the probe is characterised by the $4-$momentum transfer squared ($-Q^2$) in the process which is the virtuality of the ``exchanged'' virtual photon; the ``struck'' parton (quark $q$) has an inelasticity $x$ with respect to its parent $\gamma$; b) Feynman diagram illustrating the mechanism which can account for the interaction of an electron with a particle with hadronic structure (here a proton $4-$momentum $p$): a ``ladder'' of ``partons'', in this case the quanta of the QFT, quarks $q$ and gluons $g$ of QCD with inelasticity (fractional momenta) variables $x_{i/p}$ and $x_{i/\gamma}$, forms the basis of the interaction dynamics; the association of the partons in the ladder with the structure of the proton or with the photon is arbitrary, requiring a priori assumption concerning transverse momentum $P_T$ and and rapidity $\eta$ relative to that of the parent photon or proton.}
   \label{fig:lowxdiag}
\end{center}
\vspace{-12.6cm}

a)
\hspace{8.5cm}
b)

\hspace{10cm}\begin{large}$ep\rightarrow eX$\end{large}
\vspace{0.4cm}

\hspace{1cm}\begin{large}$e\gamma\rightarrow eX$\hspace{7cm}$e$\end{large}
\vspace{0.2cm}

\hspace{11cm}$Q^2$ $y$\hspace{0.6cm}$q$

\begin{large}$e$\end{large}
\vspace{0.2cm}

\hspace{11.6cm}$p_q$\hspace{0.3cm}Bjorken$-x$

\hspace{0.5cm}virtual photon $\gamma^*$\hspace{0.5cm} $Q^2$
\vspace{0.2cm}

\hspace{7.5cm}$x_{i/\gamma}$ $i=q,g$ increasing\hspace{0.25cm}\begin{Large}$\uparrow$\hspace{0.3cm}\end{Large}
\vspace{0.2cm}

\hspace{9.7cm}rapidity $\eta$\hspace{0.3cm}\begin{Large}$\updownarrow$\hspace{0.3cm}\end{Large}\hspace{0.8cm}$P_T$\hspace{0.3cm}\begin{Large}$\leftrightarrow$\hspace{0.3cm}\end{Large}

\hspace{1.7cm}struck quark\hspace{0.5cm}Bjorken$-x$ 

\hspace{7.5cm}$x_{i/p}$ $i=q,g$ increasing\hspace{0.3cm}\begin{Large}$\downarrow$\hspace{0.3cm}\end{Large}
\vspace{0.1cm}

\hspace{10cm}$p^2=-P^2$
\vspace{-0.2cm}

\begin{large}$\gamma$\hspace{8.3cm}proton\end{large}
\vspace{4.4cm}

\end{figure}

A simple way to observe the effects of structure is to probe in a scattering experiment with a particle which is ``point--like''~\cite{Taylor}, for example with a fundamental quantum like the electron (figure~\ref{fig:lowxdiag}a), acknowledging the theme of this paper that such a probe will also have structure which can in certain circumstances complicate matters (see below). The target of mass squared $p^2=-P^2$ is probed at $4-$momentum transfer squared scale $q^2=-Q^2$ which is appropriate to resolve any structure in it and which inevitably causes the break--up, that is fragmentation, of it. Such experiments are known collectively as ``deep--inelastic scattering'' or DIS.

The well established formalism of ``structure functions'' is used, into which all unknowns associated with the putative structure are lumped together (the ``blob'' in figure~\ref{fig:lowxdiag}a). The cross--section is written\footnote{\label{foot:DIS}A framework for deep--inelastic scattering of an electron off a target is assumed in which only virtual photon, and not massive $Z$ and $W$, exchange is considered. As written equation~\ref{eq:structurefunctions} is also applicable only to a spin$-\frac{1}{2}$ point--like probe interacting with an unpolarised target with $2$ spin degrees of freedom (proton with helicity $\pm 1/2$, real photon with helicity $\pm 1$). For virtual photons with helicity $\pm 1,0$ and space--like mass squared things are a little more complicated~\cite{Maxfield,Budnev,Rossi}.}~\cite{Taylor,Foster,Sciulli}
\begin{equation}
\frac{{\rm d}^2\sigma}{{\rm d}x{\rm d}Q^2}=\frac{4\pi\alpha^2}{xQ^4}\{1-y+\frac{y^2}{2[1+R(x,Q^2)]}\}F_{2}(x,Q^2)
\label{eq:structurefunctions}
\end{equation}
in terms of two structure functions $F_2(x,Q^2)$ and $R(x,Q^2)$. $R(x,Q^2)$ is strictly speaking the ratio of a second structure function to $F_2(x,Q^2)$. Two are necessary because the exchanged photon (figure~\ref{fig:lowxdiag}a) is off--shell and can therefore be either transverse (T, helicity $\pm 1$) or longitudinal (L, helicity $0$). The cross--section $\frac{{\rm d}^2\sigma}{{\rm d}x{\rm d}Q^2}$ is specified almost entirely by $F_2$ unless measurements are made at large $y$. 
In equation~\ref{eq:structurefunctions} $y$ is the inelasticity (equation~\ref{eq:y}) and $Q^2$ is the virtuality (equation~\ref{eq:tdef}) of the QED splitting at the electron--photon vertex. In terms of the energies $E$ and $E'$ and direction (polar angle $\theta$) of the incident and scattered (massless) electrons in the laboratory frame, which does not approximate to the ``infinite momentum frame'', $y$ and $Q^2$ are given by
$$y=1-\frac{E'+E'\cos\theta}{E}=1-\frac{E'}{2E}\cos^2\frac{\theta}{2}$$ and
\begin{equation}
Q^2=-q^2=4EE'\sin^2\frac{\theta}{2}\,\,\,\,\stackrel{\theta{\rm small}}{\longrightarrow}\,\,\,\, \frac{p^2_T}{1-y}.
\label{eq:Q2}
\end{equation}

The variable $x$ (Bjorken$-x$) is the inelasticity of the ``struck'' parton with respect to its parent, by definition a quark because it must couple to the photon (figure~\ref{fig:lowxdiag}a). From the relativistically covariant formulation of inelasticity (equation~\ref{eq:y}) and its interpretation in the infinite momentum frame of the parent (equation~\ref{eq:yinfmomentum}), it is identified as the fraction of the parent (target) momentum carried by the struck parton in this frame. The struck parton is assumed to have small virtuality $|t|$ in the target splitting. It can be expressed in terms of invariants as follows (following the notation in figure~\ref{fig:lowxdiag}b),
\begin{equation}
x=\frac{p_q.q}{p.q}=\frac{Q^2-t}{2q.p}\approx\frac{Q^2}{Q^2+W^2+P^2}
\label{eq:xBj}
\end{equation}
where $W$ specifies the invariant mass of the system of final state particles other than the scattered point--like probe $e$, and the mass squared of the target is $p^2=-P^2$. 

If we assign $f_{i/T}(x,Q^2)$ to describe the probability of finding a quark of species $i$ with inelasticity $x$ in the target $T$ when probed at scale $Q^2$, then comparison of equation~\ref{eq:structurefunctions} with the expression for the elastic scattering of $e$ with a spin$-\frac{1}{2}$ quark requires
\begin{equation}
F_2(x,Q^2)=\sum_i e_i^2xf_{i/T}(x,Q^2).
\label{eq:xf}
\end{equation}
Thus, in the ``infinite momentum frame'' the structure function $F_2$ can be thought of as the ``momentum weighted'' probability of finding the partons (quarks with charges $e_i$) in the target. The $f_{i/T}$ describe the splitting of the probed structure into remnant and one parton $i$. They are the quantities which specify precisely the structure of the target. In general one could also take the $f_{i/T}$ to be functions of the virtuality $|t|$ of the partons, but to good approximation this is not necessary if the structure being probed is that of a hadron. Only if the target is point--like and itself a parton, so that any structure arises from quantum fluctuation, is it possible to contemplate calculating the $f_{i/T}$, and thereby $F_2$, in a perturbative manner in terms of the ``splitting functions'' ${\cal P}_{i/T}$ of the appropriate QFT. Then the dependence on virtuality $|t|$ of the partons must be included (see below). In the case of hadronic structure due to fluctuation (figures~\ref{fig:bubble}b, \ref{fig:bubble}c and \ref{fig:bubble}d), the calculations of course use QCD. In general the $xf_{i/T}$ are referred to as ``parton density functions'' (pdf).

\subsubsection{Structure, Dynamics, and Factorisation}
However, following Gribov et al.~\cite{Gribov}, any further consideration of the nature of a high energy interaction leads quickly to the realisation that the distinction between what is interpreted as structure and what is interpreted as interaction dynamics is not clear cut. This is obvious even in the case where one of the interacting particles is point--like, such as in $ep$ DIS, if one considers any sort of realistic Feynman diagram as in figure~\ref{fig:lowxdiag}b). The possibility of a complicated interaction mechanism involving many field quanta amply demonstrates that there is no unique association of structure either with the fundamental, and therefore point--like, photon or with the target with ``valence--structure''. Where one associates emitted ``partons'', that is the QCD quark $q$ and gluon $g$ field quanta, with the structure of either the photon or the target, or with the consequences of the interaction dynamics, is arbitrary with no further a priori assumptions. This does not mean that the formalism in equation~\ref{eq:structurefunctions} is incorrect, only that the interpretation of the results encapsulated in $F_2$ as those describing a universal proton structure may well run into difficulties without further definition.

Thus in practice in any high energy interaction, the distinction between structure and interaction is at a quoted ``factorisation scale''. This is a momentum transfer squared scale below which any parton activity is considered to be part of the parent structure and above which any parton activity is considered to be part of the interaction dynamics. Thus the factorisation scale corresponds in any process to the scale at which structure is probed. In $ep$ DIS (figure~\ref{fig:lowxdiag}), the factorisation scale is clearly $Q^2$ --  this is reasonable because $Q^2$ is almost always the largest momentum transfer squared scale in the DIS interaction, and so it is self--evident that all ``hadron activity'' at lower momentum transfer should be associated with target structure. It is with this in mind that one resolves the conundrum in the first sentence of this section.

There is however the possibility that interactions take place in which some partons are produced with a larger transverse momentum squared $P_T^2$ than $Q^2$. There will thereby be partons of lower $P_T^2$ which have rapidities closer to either the photon or the target (figure~\ref{fig:lowxdiag}b) and which therefore carry larger fractions $x_{i/\gamma}$ and $x_{i/p}$ of the interacting photon and proton momenta respectively\footnote{Here the infinite momentum frame view is implicit, and additional covariant ``fractional momentum'' variables $x_{i/T}=\frac{p_i.q}{p.q}$ and $x_{i/\gamma}=\frac{p_i.p}{q.p}$ are introduced specifying the momentum of parton $i$ as a fraction of the momentum of the parents $T$ and $\gamma$ when the latter have large momenta.} (equation~\ref{eq:ytorapidity}).
These are naturally associated with structure in the probe photon and the target respectively (figure~\ref{fig:lowxdiag}), and the factorisation scale is then taken to be $P_T^2$. 

Any structure function is thus dependent on the choice of factorisation scale. It is found that structure functions for hadrons with intrinsic spatial extent, which are specified in terms of the $x-$dependences of each parton species (equation~\ref{eq:xf}) at a chosen momentum transfer scale for factorisation, can be used universally in any high energy interaction at that scale~\cite{Stirling}. In the same way, and with the picture of point--like structure in QFT described above in mind, it is also to be expected that photon structure will be similarly universal and factorisable.

\section{Photon Structure}
\subsection{QED Photon Structure}
\begin{figure}[hbt]
  \begin{center}
\hspace*{0.4cm}\epsfig{file=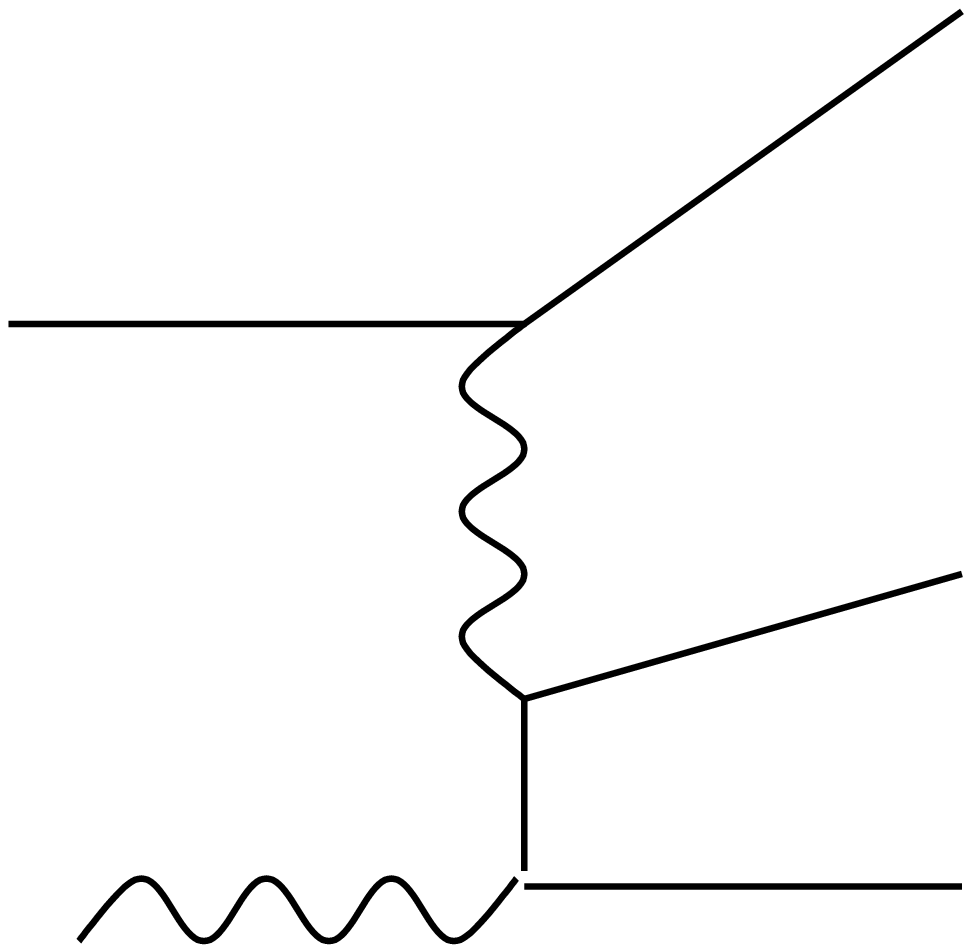,width=0.4\textwidth}\hfill\epsfig{file=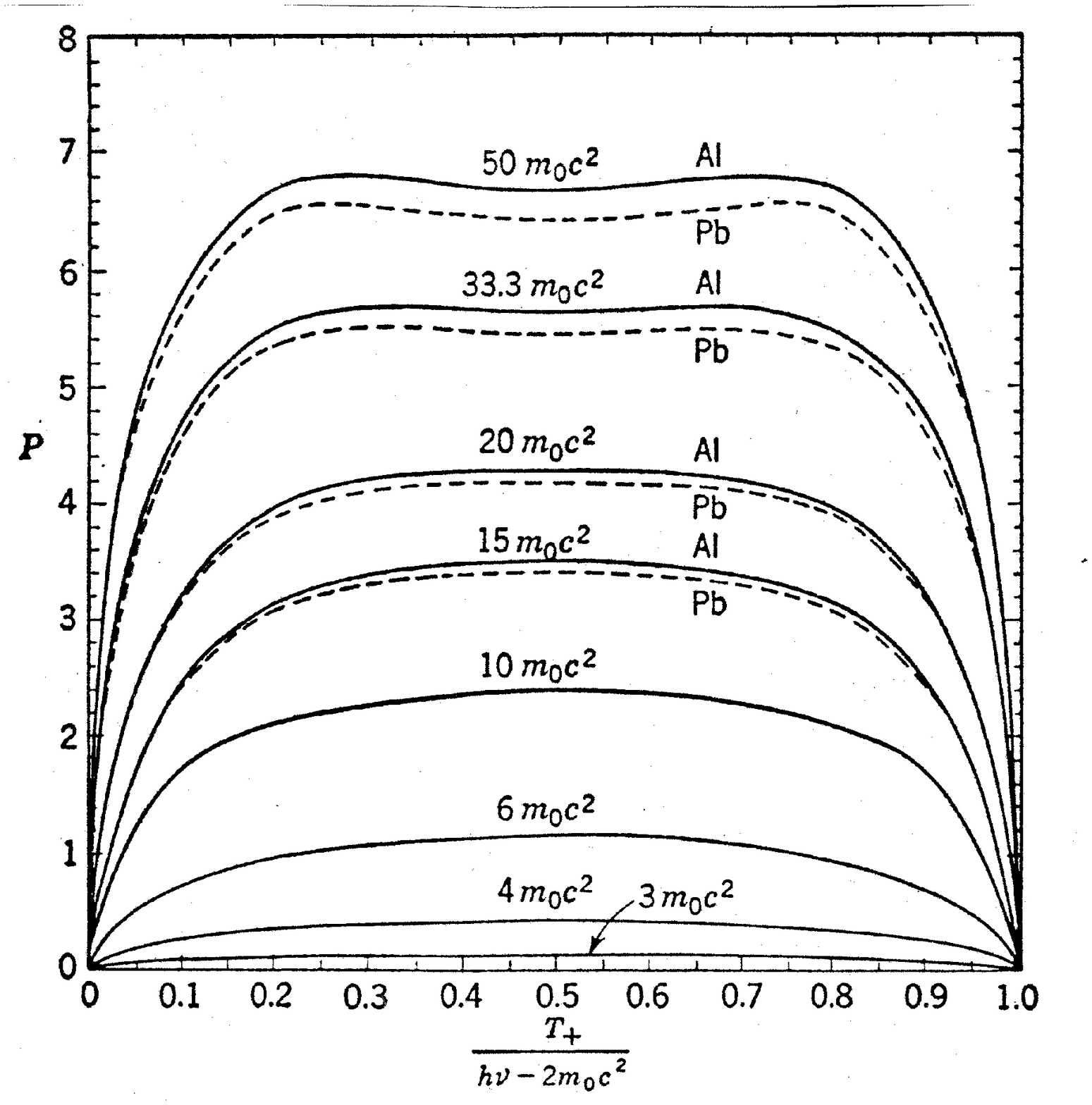,width=0.49\textwidth}
\caption{\footnotesize a) Diagram illustrating $e^+e^-$ pair conversion of an $X-$ray photon of energy greater than threshold $2m_oc^2\sim 1$ MeV ($m_0$ is the electron mass); the splitting of the photon is probed by the field of the nearby atomic nucleus such that the off--shell $e^{\pm}$ with space--like mass squared $t$ absorbs energy and emerges on--shell together with the $e^{\mp}$; b) calculation of the splitting function ${\cal P}_{e/\gamma}$ where the inelasticity variable $x=\frac{T_+}{E_{\gamma}-2m_o}$ is here taken as the ratio of the kinetic energy of the positron to the total available; ${\cal P}_{e/\gamma}$ is slowly varying between its limiting values of zero at $x=0$ and $x=1$ also visible is the slow (logarithmic) dependence of ${\cal P}_{e/\gamma}$ on $X-$ray energy; the slightly different results (continuous and dashed) for the different nuclei Al and Pb are due to the different atomic electron configurations of each.}
   \label{fig:Aprobegamma}
\end{center}
\vspace{-12.4cm}

a)\hspace{7.5cm}b)
\vspace{2.1cm}

\begin{large}$A$\end{large}
\vspace{1cm}

\hspace{0.4cm}$P^2_T\,\gapprox\, 1$ MeV

\hspace{7.1cm}\begin{large}$e^{\pm}$\end{large}
\vspace{0.8cm}

\hspace{1.6cm}$x=\frac{T_+}{E_{\gamma}-2m_e}$\hspace{0.6cm}$t$
\vspace{0.2cm}

\hspace{0.4cm}\begin{large}$\gamma$\hspace{6.5cm}$e^{\mp}$\end{large}
\vspace{4.6cm}

\end{figure}

The first measurements which probed the point--like structure of the photon can be traced back to experiments at the time of the discovery of the positron. By making measurements of the absorption of $X-$rays in matter, Tarrant~\cite{Tarrant} and Chao~\cite{Chao} first demonstrated anomalously high absorption for $X-$rays with energy above about $1$ MeV. There followed the classic experiment of Anderson and Neddermeyer~\cite{AndersonNeddermeyer} which demonstrated that this was attributable to the conversion of the $X-$rays into positrons $e^+$ and electrons $e^-$. Subsequently Oppenheimer and Plesset~\cite{OppenheimerPlesset} showed that the results were consistent with a process in which the $X-$ray photon split into $e^+e^-$ in the electromagnetic field of the atoms in the absorber. Armed with the first formulation of QED, Bethe and Heitler then made the first precision calculation of the cross--section~\cite{BetheHeitler}. Figure~\ref{fig:Aprobegamma}a) illustrates the process, and demonstrates how it can thus be thought of as the nucleus $A$ probing the QED structure due to fluctuation of the $X-$ray photon.

The most notable features of the cross--section for the $e^+e^-$ pair production process are that the dependence of the cross--section, through the splitting function ${\cal P}_{e/\gamma}$, on the inelasticity variable $x=\frac{T_+}{E_{\gamma}-2m_e}$, is slowly varying (figure~\ref{fig:Aprobegamma}b), and that there is a slow logarithmic dependence on $X-$ray energy, namely
\begin{equation}
{\cal P}_{e/\gamma}(x,W^2_{ee})= e^4\frac{\alpha}{\pi}\{[x^2+(1-x)^2]\ln\frac{W^2_{ee}}{m_e^2}+8x(1-x)-1\}
\label{eq:splitting}
\end{equation}
where $e$ and (here) $m_e$ are the charge and mass of the electron, $W_{ee}$ is the invariant mass of the $e^+e^-$ system, and $\alpha$ is the fine structure constant. The logarithmic dependence arises from the need when calculating ${\cal P}_{e/\gamma}$ to integrate the splitting probability (proportional to $1/(m_e^2-t)$; section~\ref{sub:pointlikestructure}) over the space--like mass squared $t$ of the off--shell electron/positron in the photon splitting (figure~\ref{fig:Aprobegamma}a). In the measurement the inelasticity $x$ of the positron constituent is determined directly from the ratio of kinetic energy of the positron to the total available kinetic energy, that is the laboratory is to good approximation the ``infinite momentum frame'' of the parent photon -- equation~\ref{eq:yinfmomentum}. 
In terms of the nucleus probing the QED structure of the photon, the probe scale is $P^2_T\,\gapprox\, 1$ MeV$^2$ which amounts to spatial resolution of $\sim 200$ fm. This is many times the size of the nucleus, which therefore acts like a point--like probe of QED on a scale of atomic dimension ($1$ \AA $\,=10^5$ fm). Using equation~\ref{eq:transversesize} and remembering that the cross section integrates over the space--like mass squared $t$ of the off--shell electron, the dimension of the structure of the photon is characterised by a scale which is the mass of the electron or less, namely $\sim 0.5$ MeV or $\sim 400$ fm, so that the QED structure of the photon is not here probed in great detail.

Similarly beautiful measurements continue to this day to probe QED pair conversion, and thus the QED structure of the photon, at much shorter distances. In high energy $e^+e^-$ colliders interactions of the type $ee\rightarrow ee\mu^+\mu^-$, in which an electron or positron are scattered with substantial $Q^2$ and thereby probe the remainder of the interaction, can be used to probe photon conversion to two muons. The electron or positron which scatters with high $Q^2$ replaces the atomic nucleus as probe in figure~\ref{fig:Aprobegamma}a). The ``target'' photon is constrained to be almost real (virtuality $P^2\sim 0$) by requiring the second electron/positron to scatter through very small angles (equation~\ref{eq:Q2}), usually such that it stays undetected inside the beam pipe of the $e^+e^-$ collider. The data are analysed in terms of a structure function $F_{2\,QED}^{\gamma}$ (equation~\ref{eq:structurefunctions} for $e\gamma\rightarrow e\mu^+\mu^-$), for which the expressions ($m_e$ is here the electron rest mass)
\begin{equation}
f^{\rm T}_{\gamma/e}(y,P^2) = 
\frac{\alpha}{2\pi}\Bigl[\frac{1+(1-y)^2}{y}\frac{1}{P^2}-\frac{2m_e^2y}{P^4}\Bigr]\,\,\,\,\,\,\,\,\,\,\,\,\,\,\,\,\,\,\,f^{\rm L}_{\gamma/e}(y,P^2) = 
\frac{\alpha}{2\pi}\frac{2(1-y)}{y}\frac{1}{P^2}
\label{eq:photonfluxes}
\end{equation}
due to electron splitting $e\rightarrow e\gamma^*$ in QED (figure~\ref{fig:bubble}d) with quark replaced by electron and gluon replaced by photon), or close approximations to them, are used to specify the target flux of photons~\cite{Budnev}. The struck muon inelasticity is here Bjorken$-x$ which is determined this time from an accurate measurement of the $\mu^+\mu^-$ invariant mass $W_{\mu\mu}$ and equation~\ref{eq:xBj} with $W=W_{\mu\mu}$. 

\begin{figure}[hbt]
  \begin{center}
\epsfig{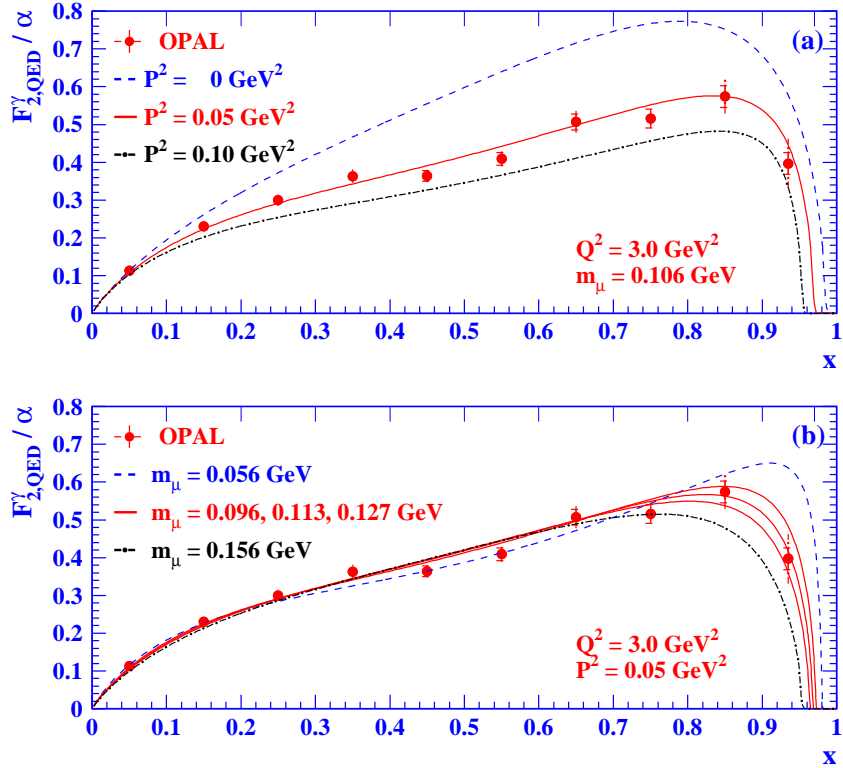}
\caption{\footnotesize Measurement of the structure function $F_{2\,QED}^{\gamma}$ of the photon in the QED process $e\gamma\rightarrow e\mu^+\mu^-$ by the OPAL experiment at LEP; the curves show the expectation for different values of the muon mass $m_{\mu}$ and different (small) virtualities of the target photon $P^2$; the characteristic rising dependence on $x$ arising from the QED splitting $\gamma\rightarrow \mu^+\mu^-$ is obvious, similar to that at much longer distance scale in $X-$ray pair production (figure~\ref{fig:Aprobegamma}).}
   \label{fig:OPALQEDF2}
\end{center}
\end{figure}

Theoretical expectation requires only one flavour of ``constituent'', the $\mu$, whose pdf $f_{\mu/\gamma}(x,Q^2)$ is calculable with the QED splitting function ${\cal P}_{\mu/\gamma}$ for $\gamma\rightarrow\mu^+\mu^-$ (equation~\ref{eq:splitting}). Thus the momentum weighted splitting function $x{\cal P}_{\mu/\gamma}=xf_{\mu/\gamma}$, and using equation~\ref{eq:xf} $F_{2\,QED}^{\gamma}$ is expected to rise approximately linearly away from the extreme values of $x$ according to
$$F_{2\,QED}^{\gamma}=xf_{\mu/\gamma}(x,Q^2)=x{\cal P}_{\mu/\gamma}(x,Q^2)=e^4\frac{\alpha}{\pi}\{x[(x^2+(1-x)^2]\ln\frac{W_{\mu\mu}^2}{m_{\mu}}+8x^2(1-x)-x\}.$$ 
The results~\cite{OPALQED} in figure~\ref{fig:OPALQEDF2} thereby constitute a beautiful update of the original measurements of $X-$ray pair production, but now for a QED fluctuation of the photon of dimension characterised by the mass of the muon (equation~\ref{eq:transversesize}), namely $0.1$ GeV or $\sim 2$ fm, with a probe of scale $Q^2$, that is with precision about $0.07$ fm.

\subsection{Hadronic Photon Structure}

\subsubsection{Principles}
The principles underlying the way in which the QED structure of the photon is measured above extend in a straightforward way to the hadronic structure of the photon. The process $ee\rightarrow eeX$, in which $X$ is the inclusive set of all hadronic final states kinematically accessible, can be analysed in terms of the process $e\gamma\rightarrow eX$. The $e$, which scatters with the larger $Q^2$, probes the hadronic structure of the photon $\gamma$, which couples to the other $e$ with lower $4-$momentum squared $p^2=-P^2$ (figure~\ref{fig:lowxdiag}a). Measurements of a structure function $F_2^{\gamma}(x,Q^2)$ are possible for a range of values of target virtuality $P^2$ by analysing the data in the framework of the usual structure function (equation~\ref{eq:structurefunctions} and footnote~\ref{foot:DIS}). There are some further complications of a technical nature arising because of the incomplete acceptance of experiments for the detection of all the hadronic final state of mass $W$, and thereby the kinematic reconstruction of Bjorken$-x$ with equation~\ref{eq:xBj}~\cite{PLUTOF2real,others}.

\subsubsection{The Hadronic Photon Structure Function \mbox{{\boldmath $F_2^{\gamma}$}}}
\begin{figure}[hbt]
  \begin{center}
\epsfig{file=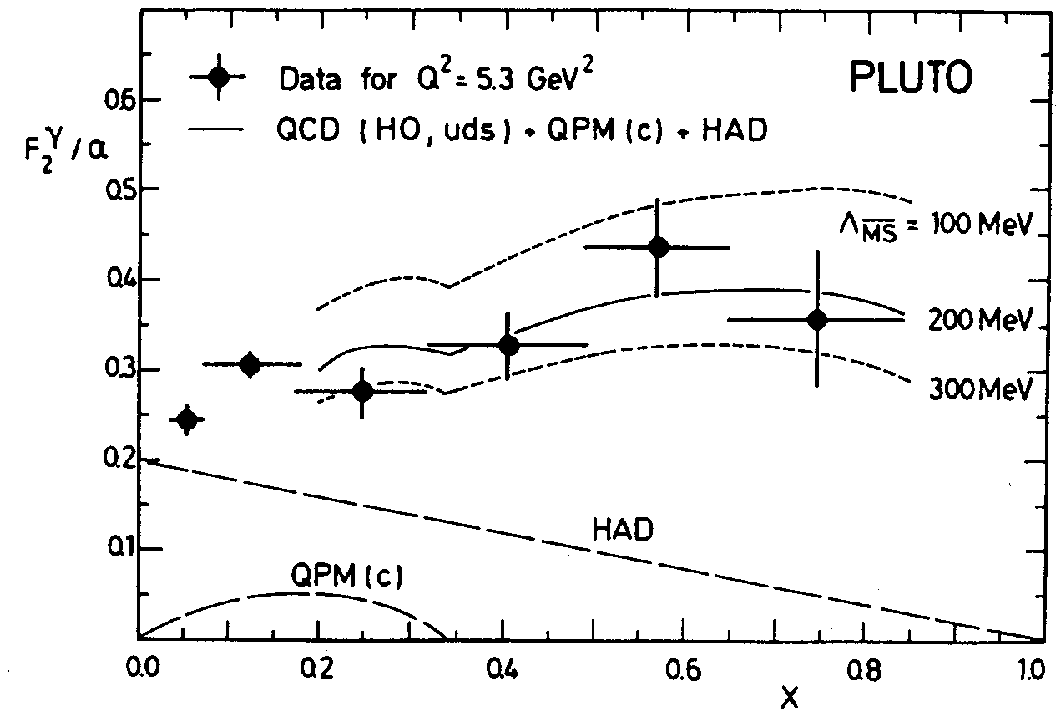,width=0.49\textwidth}\epsfig{file=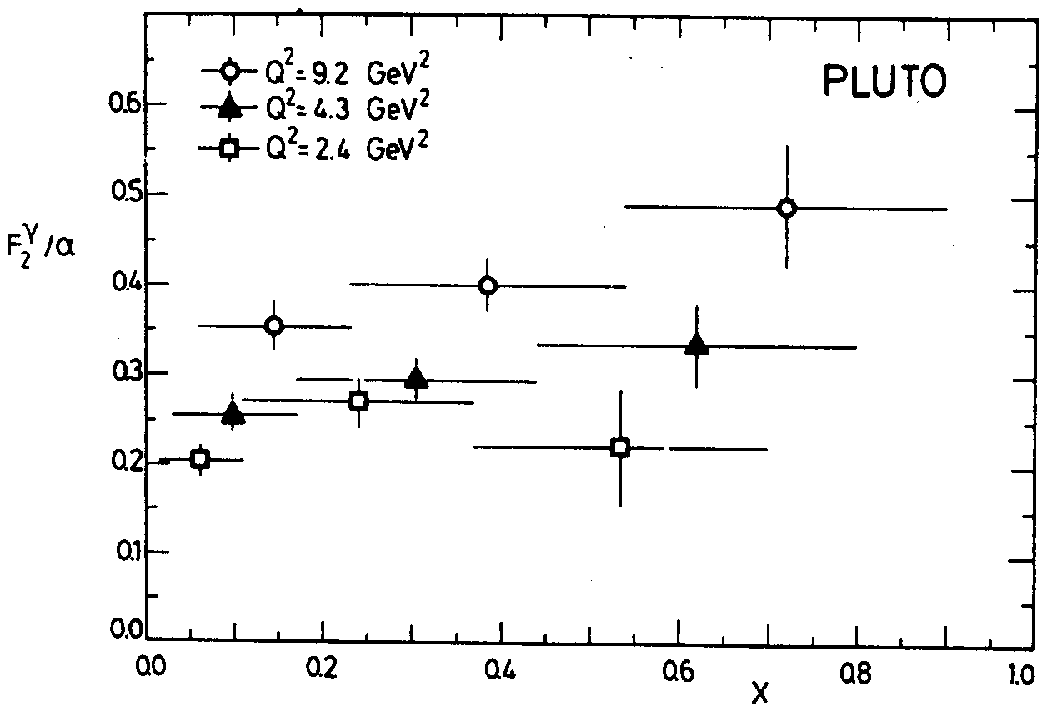,width=0.49\textwidth}
\caption{\footnotesize Measurement of the structure function $F^{\gamma}_2(x,Q^2)$ of the real photon ($P^2=0$) by the PLUTO experiment at the PETRA $e^+e^-$ storage ring in the process $e\gamma\rightarrow eX$ where $X$ is a hadronic system: a) the Bjorken$-x$ dependence at fixed $Q^2$; the contribution due to VMD is marked HAD, the contribution due to charm quark production is marked QPM(c), and the theoretical curves are labelled with different momentum transfer cut--offs $\Lambda_{\overline{\rm MS}}$; b) the Bjorken$-x$ dependence for three different $Q^2$ showing the slight (approximately logarithmic) increase with increasing $Q^2$ of $F_2^{\gamma}$.}
   \label{fig:PLUTOF2x}
\end{center}
\vspace{-9cm}

a)\hspace{8.3cm}b)
\vspace{7.8cm}

\end{figure}

The main features of measurements of hadronic photon structure are visible in early results (figure~\ref{fig:PLUTOF2x}) at the PETRA $e^+e^-$ storage ring~\cite{PLUTOF2real} of the structure function $F_{2}^{\gamma}$ of the real photon ($P^2=0$). The $x$ dependence is noteworthy in that $F_2^{\gamma}$ shows no sign of any decline with increasing $x$, in contrast to the $x$ dependence of the structure function $F_2$ of the proton~\cite{Foster,Sciulli}. There is a slowly rising dependence with increasing $Q^2$ for most of the measured $x$ range. Bearing in mind what one learns from the QED structure of the photon, both these observations are in qualitative accord with the expectation of the hadronic structure of the photon being driven by its QFT splitting into fermion anti--fermion, in this case presumably quark and anti--quark ($q{\bar q}$) in QCD.

\begin{figure}[hbt]
  \begin{center}
\epsfig{file=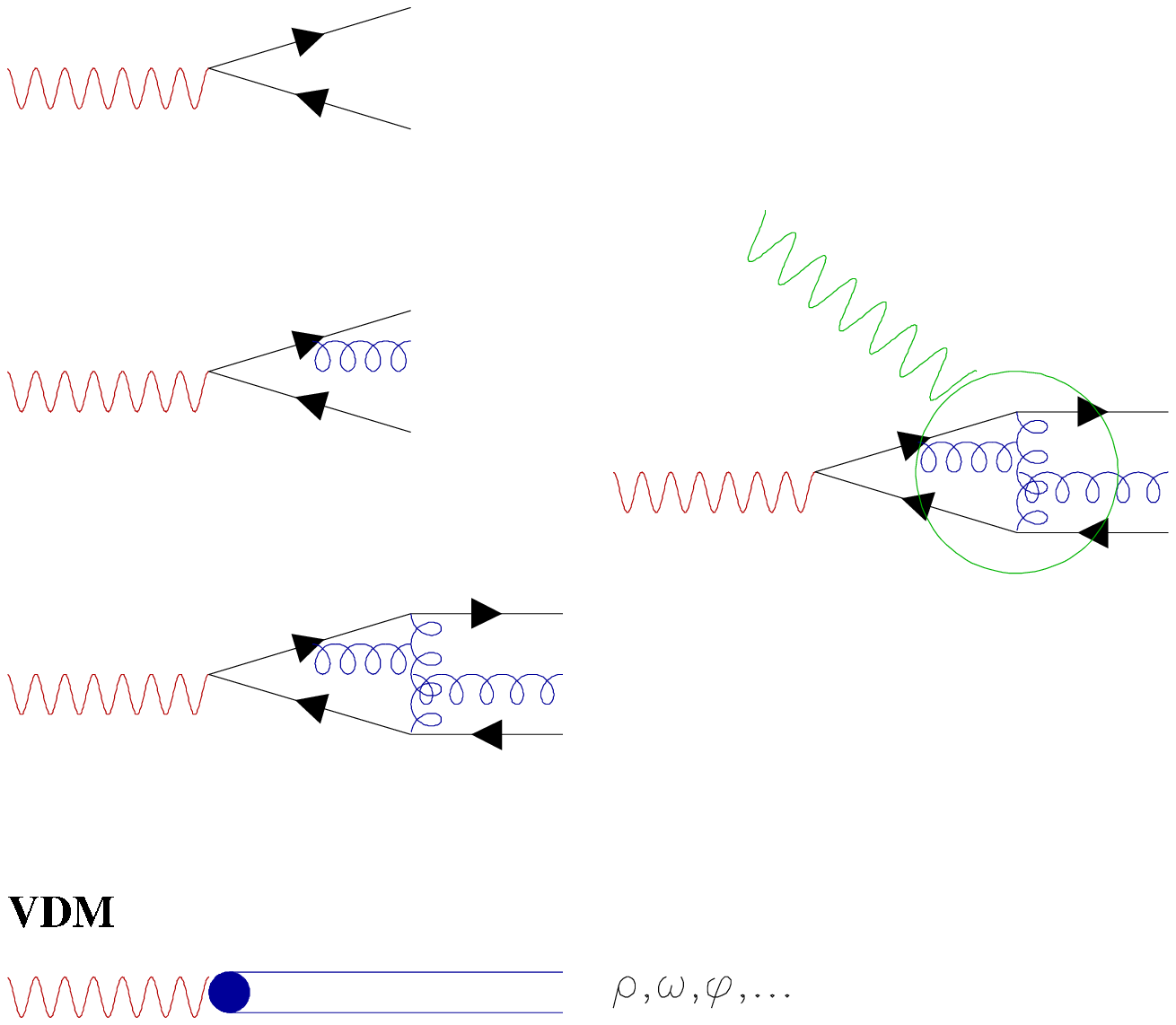,width=0.6\textwidth}
\end{center}
\vspace*{-9.85cm}

\hspace{1.5cm}\begin{small}{\bf a) point--like: calculable - LO} \mbox{{\boldmath $\propto \ln Q^2$}}
\hspace{3cm}\mbox{{\boldmath$\gamma^*\gamma^*\rightarrow q{\bar q}$}} 

\hspace{10cm} {\bf box diagram} \mbox{{\boldmath $t\,\lapprox\,-\Lambda^2$}}

\hspace{10cm} \mbox{{\boldmath $+$}}{\bf  QCD corrections} 

\hspace{10.7cm} \mbox{{\boldmath $q{\bar q}+q{\bar q}g+...\,\,$}}   
\vspace{0.4cm}

\hspace{8.8cm}\mbox{{\boldmath$Q^2$}}   
\vspace{1.3cm}

\hspace{10.3cm}\mbox{{\boldmath$t$}}

\hspace{7.6cm}\mbox{{\boldmath$P^2$}}
   
\vspace{3.1cm}

\hspace{2.4cm}{\bf b)}\hspace{1.8cm}{\bf : ``hadron'' -- incalculable} \mbox{{\boldmath $t\,\gapprox\,-\Lambda^2$}}   
\vspace{1.15cm}

\hspace{3cm} \mbox{{\boldmath $\underbrace{F_2^{\gamma}=3\sum_q e^4_q\frac{\alpha}{\pi} x[x^2+(1-x)^2]\cdot[\ln\frac{Q^2}{P^2}}$}{\boldmath$+$}{\boldmath $\underbrace{\ln\frac{P^2}{\Lambda^2}]+F_2^{{\rm VMD}}(\propto\frac{1}{P^4})}$}}   
\vspace{0.2cm}

\hspace{5cm}{\bf point--like} \mbox{{\boldmath $P^2>\Lambda^2$\hspace{2cm} $P^2\leq\Lambda^2$}}{\bf - double counting?}
\end{small}
  \begin{center}
\caption{\footnotesize Some QCD diagrams which contribute to the hadronic structure of the photon of mass squared $-P^2$; a) diagrams which can be included in a perturbative QCD calculation involving the addition of gluons to the lowest order splitting $\gamma\rightarrow q{\bar q}$ (called the ``box diagram''); these diagrams are often included together under the generic term ``point--like''; $t$ is the $4-$momentum transfer squared, that is the virtuality, of the quark in the ``box--diagram''; the calulation of the contribution of these diagrams to $F_2^{\gamma}$ requires integration of the box--diagram amplitude over $t$, and it generates the leading logarithm contribution $\propto \ln Q^2$; b) schematic diagram illustrating the extreme in which the gluon modifications to $\gamma\rightarrow q{\bar q}$ distort the hadronic structure of the photon to the extent that it assumes the properties of a bound state hadron, namely (by quantum number conservation) a vector meson -- this contribution, which is arbitrarily identified with the ``hadron'', incalculable piece in Witten's calculation (see text), is referred to generically as Vector Meson Dominance or VMD or VDM. In the final expression for $F_2^{\gamma}$, the LO QCD contribution is written as the sum of two terms such that it is clear that for $P^2>\Lambda^2$ only the pQCD box--diagram $\gamma^*\gamma^*\rightarrow q{\bar q}$ plus QCD corrections remain.}
   \label{fig:QCDdiags}
\end{center}
\end{figure}

A quantitative understanding of $F_2^{\gamma}$ and photon structure in terms of the coupling of quark and anti--quark to the photon is complicated in QCD because of the strong coupling ($\alpha_S$) in the theory. The situation is summarised in figure~\ref{fig:QCDdiags} in which are the diagrams involving both gluons and quarks which build the hadronic structure of the photon. A simple LO parametrisation in terms of Bjorken$-x$
\begin{equation}
F_2^{\gamma}=3\sum_q e^4_q\frac{\alpha}{\pi}x[x^2+(1-x)^2]\ln\frac{Q^2}{\Lambda^2}+[{\rm ``hadron''}\equiv{\rm VMD}]
\label{eq:LOQCD}
\end{equation}
can be written down in QCD, in which the factor 3 arises from a summation over quark colour and in which the $e_q$ specify the quark charges (in units of electron charge) in the summation over quark flavours. It clearly demonstrates the essential features of the QED calculation (equation~\ref{eq:splitting}). The expression is implicitly asymptotic (large logarithm) in that the logarithmic term is in $Q^2$ and not $W^2$ ($\ln W^2=\ln Q^2+\ln (\frac{1}{x}-1)-\ln P^2\rightarrow \ln Q^2$), and in that the terms in Bjorken$-x$ in equation~\ref{eq:splitting} which are not multiplied by the logarithm are dropped.
Witten~\cite{Witten} and others~\cite{LlewellynSmith} showed that these remain the leading features of $F_2^{\gamma}$ in calculations which can be completed at higher order in perturbative QCD (pQCD). In particular, the factorisation of the leading $x$ and $Q^2$ dependences in equation~\ref{eq:LOQCD}, which persist as the leading feature of higher order calculations, can be seen to arise in the photon, and not in the proton or other hadron, because of the asymptotically free running coupling constant ($\alpha_S$) in QCD ($\alpha_S\propto 1/\ln Q^2$).

The scale of the (leading) logarithmic dependence of $F_2^{\gamma}$ is taken from a cut--off $\Lambda^2$ in $4-$momentum transfer squared. When the virtuality $|t|$ of the struck quark is small ($t\rightarrow 0$), the ``box--diagram'' and its pQCD additions (figure~\ref{fig:QCDdiags}a) diverge if the quarks are massless. The integration over $t$ in the perturbative calculation of $F_2^{\gamma}$ is therefore truncated to be for $|t|>\Lambda^2$ leaving the remaining incalculable ``hadron'' piece at low $|t|$. In the absence to date of anything better to do, the ``hadron'' piece is treated phenomenologically. The view is taken that contributions with $|t|$ small ($<\Lambda^2$) correspond to long distant QCD ($\alpha_S$ large) in which the original photon splitting is so distorted by effects of multiple gluon emission and absorption that the hadronic structure of the photon has the form expected of a vector meson -- Vector Meson Dominance (VMD)~\cite{VMD} or the Vector Dominance Model (VDM). In other words, for $|t|<\Lambda^2$ the hadronic structure of the photon is due to a soft, that is long range ($\sim 1$ fm), QCD fluctuation into vector mesons, the bound states of valence quark and anti--quark (figure~\ref{fig:QCDdiags}b)~\cite{FieldKapusta}.
\vspace{0.3cm}

The curves shown in figure~\ref{fig:PLUTOF2x} are obtained using a combination of a VMD piece (HAD), which is estimated from the expected structure functions of the vector mesons $\rho$, $\omega$ and $\phi$ and which is assumed to exhibit the falling Bjorken$-x$ dependence characteristic of valence--driven hadron structure such as the proton~\cite{Taylor,Sciulli}, added to a NLO pQCD calculation of the ``point--like pieces'' characterised by the cut--off $\Lambda^2_{\overline {\rm MS}}$ in box--diagram momentum transfer squared. Such an addition is fraught with ambiguity -- what is left out in the pQCD calculation and how can we be sure that it is added into the ``VMD piece''? Or alternatively what ``double counting'' is there when both pieces are included? No-one to this day knows the complete answer. Nevertheless, the complementary way that the VMD contribution preferentially fills out the low $x$ region, flattening the otherwise rising $x-$dependence of the pQCD contribution, is phenomenologically appealing and quantitatively successful.

\begin{figure}[hbt]
  \begin{center}
\epsfig{file=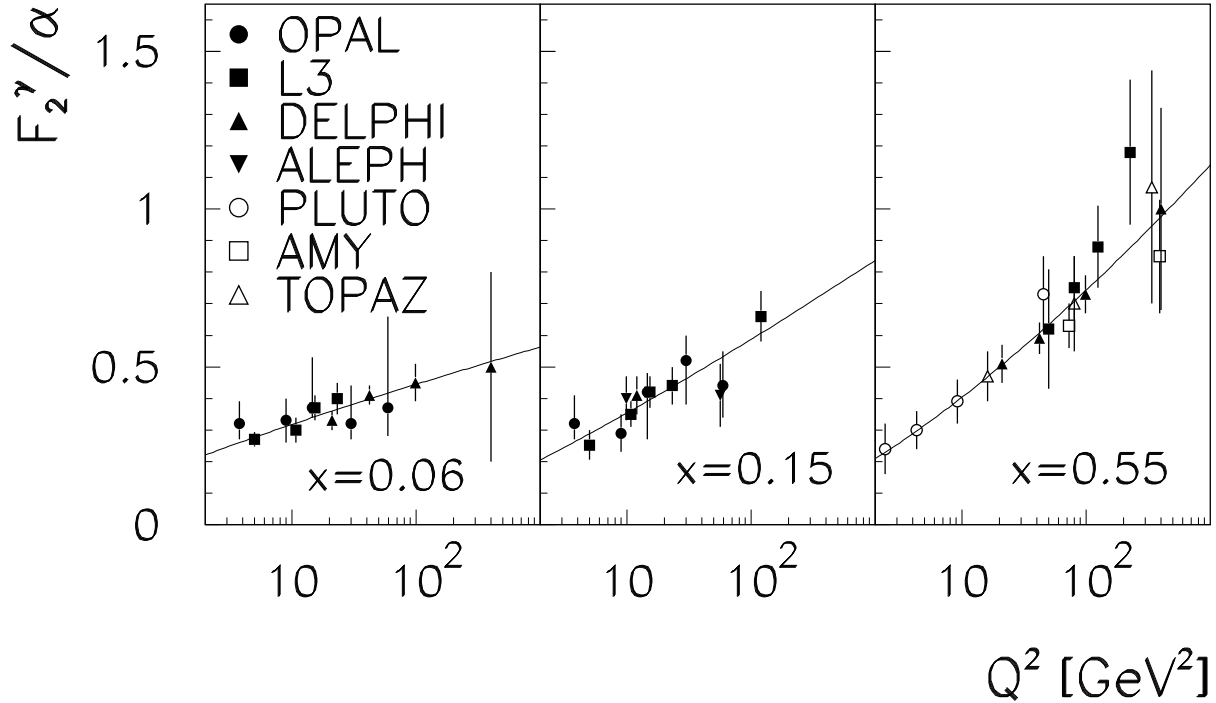,width=\textwidth}
\caption{\footnotesize The $Q^2$ dependence of the photon structure function $F_2^{\gamma}$ showing a persistent increase with increasing $Q^2$ for all Bjorken$-x$; measurements are from experiments at PETRA (PLUTO), TRISTAN (AMY, TOPAZ) and LEP (OPAL, L3,DELPHI, ALEPH); the curves are from a fit of Erdmann~\cite{Erdmann}.} \label{fig:scaling}
\end{center}
\end{figure}

Perhaps the most dramatic way in which the fluctuation--driven, structure due to $\gamma\rightarrow q{\bar q}$ and the box--diagram shines through the complications of the strong coupling of gluons in QCD is to be found in the striking ``scaling violations'' of the hadronic structure of the photon. Figure~\ref{fig:scaling} summarises~\cite{PLUTOF2real,Erdmann,OPAL,L3,DELPHI,AMY,TOPAZ,ALEPH} the $Q^2$ dependence of $F_2^{\gamma}$ for the real photon -- violations of scale invariance (scale invariance means no dependence on $Q^2$) are clear and positive in that $F_2^{\gamma}$ increases with increasing $Q^2$. The contrast with a hadron with a valence--driven quark structure is stark. For Bjorken$-x\,\gapprox\,0.15$ the structure function $F_2$ of the proton decreases with increasing $Q^2$ following the effects of the possibility of gluon Bremsstrahlung (figure~\ref{fig:bubble}d) from the valence quarks~\cite{Sciulli}. A rising dependence of $F_2^{\gamma}$ with increasing probe scale $Q^2$ throughout most of the range of $x$ is thereby seen to be {\em the} characteristic signature of fluctuation--driven (rather than ``valence driven'') structure.

Nowadays, as figure~\ref{fig:scaling} shows, measurements of $F_2^{\gamma}$ span a wide range of $Q^2$. It is therefore possible to take a more generic view of the application of QCD to photon structure much along the lines of the fits based on the DGLAP pQCD formalism~\cite{DGLAP} which are applied to measurements of the proton structure function~\cite{Foster,Sciulli}, acknowledging the differences in the fluctuation--driven, rather than valence--driven, structure of the photon. As a result, pdfs are available with which to build calculations of hard processes involving photons (see section~\ref{sec:photoproduction}), following the well tried and tested QCD approach to hard hadron interactions based on the factorisability of hadronic structure~\cite{gammapdfs,GRVLO,GRVHO}.

\subsubsection{Virtual Photon Structure}
\label{sec:virtualphotonstructure}
The picture above of photon structure can be further tested by exploiting the experimental control possible of the size, that is virtuality $P^2$, of the target photon (figure~\ref{fig:lowxdiag}a and section~\ref{sub:pointlikestructure}). In the process $ee\rightarrow eeX$ both the ``probe'' $e$ specifying $Q^2$ and the target $e$ specifying $P^2\ll Q^2$ can be detected and the energy and angle of each electron measured. The evolution of the picture above can be followed as one ``squeezes'' the photon target, that is increases $P^2$ -- working in the ``infinite momentum frame'' of the probe photon, equations~\ref{eq:ttoPTlightcone} and \ref{eq:size} tell us that increasing $P^2$ inevitably increases $|t|$, and thereby decreases the spatial extent of the structure of the photon.

Reducing the size of the photon is expected to reduce the contribution to photon hadronic structure due to VMD. The light quark (up $u$, down $d$, and strange $s$) vector mesons, $\rho(770)$, $\omega(780)$ and $\phi(1019)$, dominate this contribution. They have a dimension, characteristic of all hadrons, of about $1$ fm so that increasing $P^2$ means that there is less overlap between the (now increasingly virtual) shrinking photon and the meson wavefunctions. In $4-$momentum space, the minimum virtuality $|t|=|t_{\rm min}|$ of the off--shell quark in the fluctuation $\gamma^*\rightarrow q{\bar q}$ (figure~\ref{fig:bubble}a) increases with increasing $P^2$ (equation~\ref{eq:ttoPTlightcone} with $p_T=0$) so that because $t\neq 0$ always there is no longer any divergence in the integration over it. Once $|t_{\rm min}|$ exceeds $\Lambda^2$, the logarithmic term in equation~\ref{eq:LOQCD} no longer needs a cut--off $\Lambda$ below which the VMD contributes. Instead kinematics provides its own cut--off $|t_{\rm min}|=P^2$ which, as long as it exceeds $\Lambda^2$, means that the ``hadron'' piece in equation~\ref{eq:LOQCD} no longer exists and pQCD provides in principle a complete, and rather simple prediction for $F_2^{\gamma^*}$
\begin{equation}
F_2^{\gamma^*}=3\sum_q e^4_q\frac{\alpha}{\pi} x[x^2+(1-x)^2]\cdot\ln\frac{Q^2}{P^2}
\label{eq:virtualstructurefunction}
\end{equation}
at leading order~\cite{Uematsu}. 

\begin{figure}[hbt]
\begin{center}
\epsfig{file=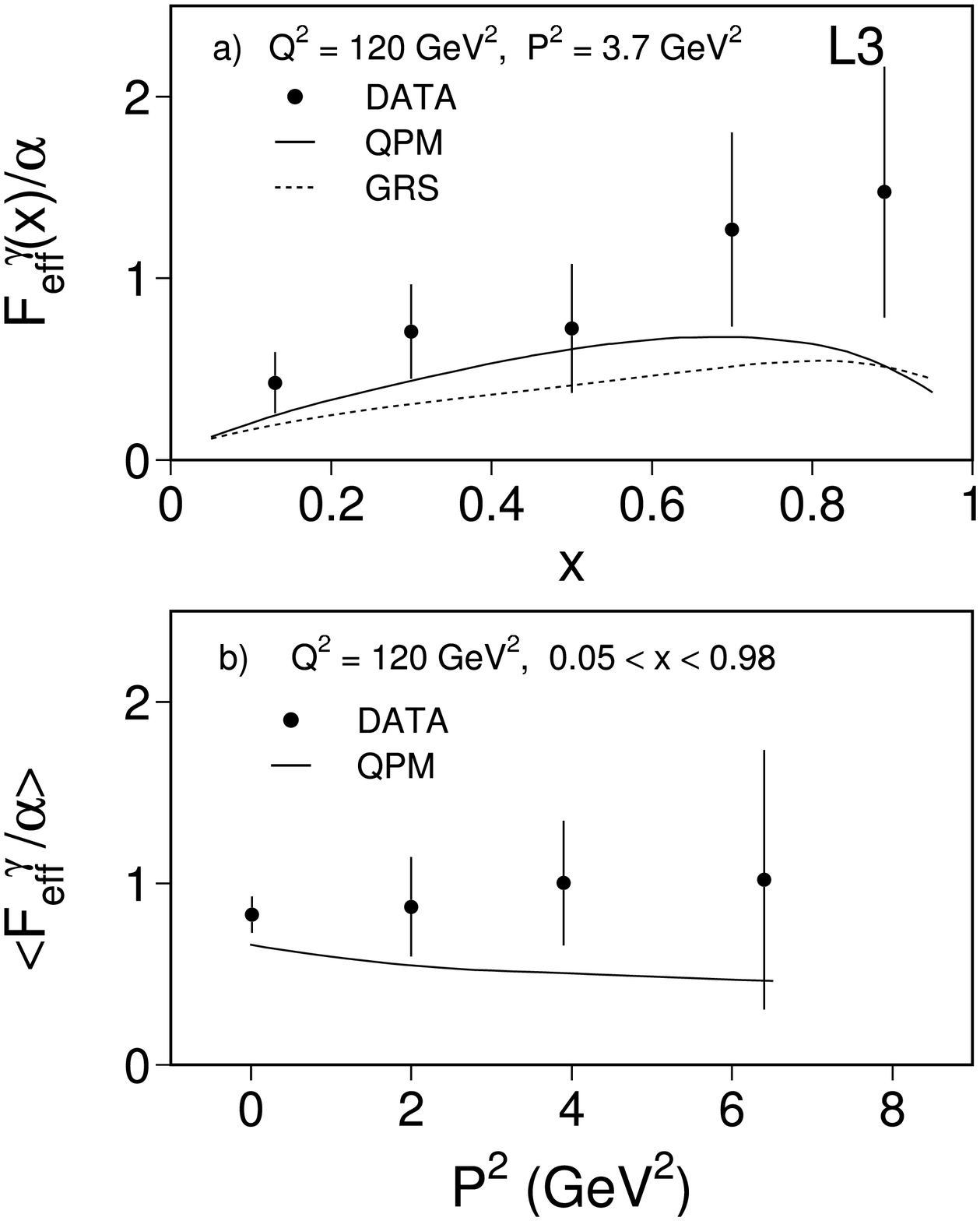,width=0.6\textwidth}
\vspace*{-12.7cm}

\epsfig{file=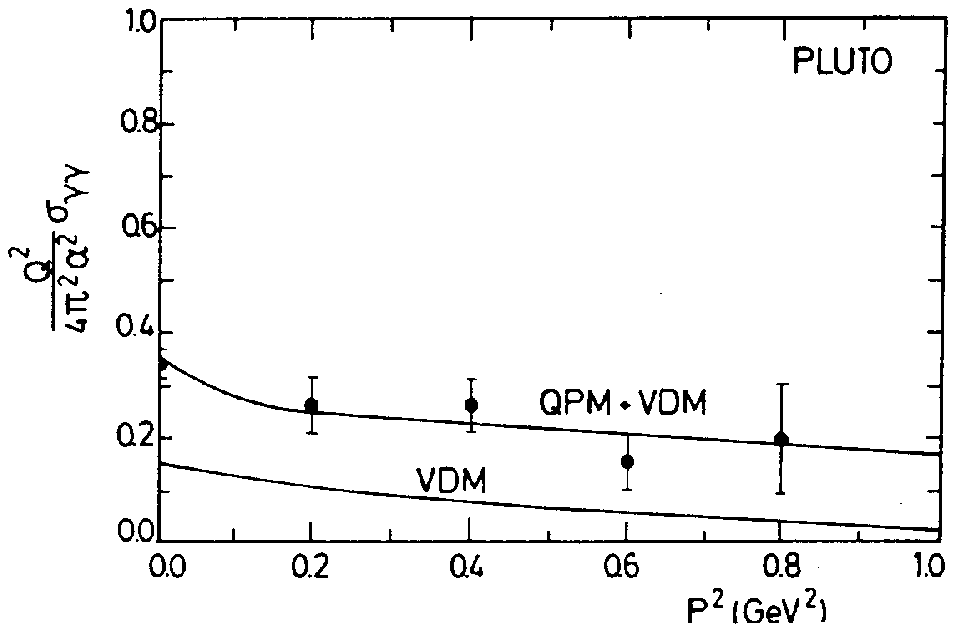,width=0.6\textwidth}\vspace*{-5.6cm}

\hspace*{0.3cm}{\bf a)} \mbox{\boldmath{$Q^2=5\,{\rm GeV^2}$}}, \mbox{\boldmath{$0.3<x<0.7$}}
\vspace*{6cm}

\hspace*{8cm}\begin{large}{\bf L3}\end{large}
\vspace*{4.8cm}

\caption{\footnotesize Dependence of the combination of photon structure functions measured in $ee\rightarrow eeX$ where both photons are off--shell on photon mass$^2$ $P^2$; a) the first results at PETRA and b) new results from LEP at higher $Q^2$ and $P^2$; QPM refers to the Quark Parton Model in which the simple QED--like $\gamma\rightarrow q{\bar q}$ box diagram is calculated for fractionally charged quarks; the curve marked VDM is an estimate of the VDM contribution.}
   \label{fig:P2dependences}
\end{center}
\end{figure}

The results, which are ultimately always statistically limited by $e^+e^-$ luminosity because the $ee\rightarrow eeX$ cross--section always falls steeply with increasing $P^2$ and $Q^2$, are shown in figure~\ref{fig:P2dependences}~\cite{L3,PLUTOvirt}. Apart from an initial decrease from $P^2=0$, there seems to be little dependence on $P^2$. The growth of the measured combination of structure functions with $Q^2$ is as expected in the factorisable dependence in equation~\ref{eq:virtualstructurefunction}
\vspace{1.2cm}

The degree to which the results in figure~\ref{fig:P2dependences} are described, or not, by the curves labelled QPM summarise how well the features discussed above are consistent with observation. The measurements are of limited precision, but it is still possible to conclude that, in terms of the VMD parametrisation assumed (VDM), there is little or no requirement for VMD for $P^2\,\gapprox\,1\,{\rm GeV^2}$ so that perturbative QCD then provides a description of photon structure. The results extend to values of $P^2\sim 6\,{\rm GeV}^2$, which means (equation~\ref{eq:transversesize} with $m_e^2=0$) that the size of the virtual photon is no larger than about $\frac{1}{P}\sim 0.08$ fm.

It is interesting here to note that the evolution of photon structure with increasing $P^2$ presumably corresponds to the evolution with increasing $Q^2$ of the dipole involved in the picture of low$-x$ electron--proton DIS discussed recently by Foster~\cite{Foster}. For in perturbative QCD, both can be thought of in terms of the evolution with decreasing transverse dimension of a $q{\bar q}$ colour dipole.

\subsubsection{Summary}
Following in the success of $ep$ DIS and the picture which it gives of proton structure, measurements of $e\gamma$ DIS reveal a remarkably clear picture of the hadronic structure of the photon. It is driven by the possibility of fluctuation into quark and anti--quark. The complications due to the strong coupling in QCD do not obscure this picture. These complications can be controlled very neatly by varying the size, that is $P^2$, of the photon, thereby also demonstrating how measurements which simultaneously specify two momentum transfer scales, $Q^2$ and $P^2$, are more amenable to contemporary QCD predictions. A complete calculation in QCD of the hadronic structure function $F_2^{\gamma}(x,Q^2,P^2)$ remains one of the essential prerequisites before QCD can truly be said to be the means of predicting hadronic phenomena.

\subsection{Photon Structure in Hadronic Photoproduction}
\label{sec:photoproduction}
\subsubsection{Principles}
\begin{figure}[hbt]
\begin{center}
\epsfig{file=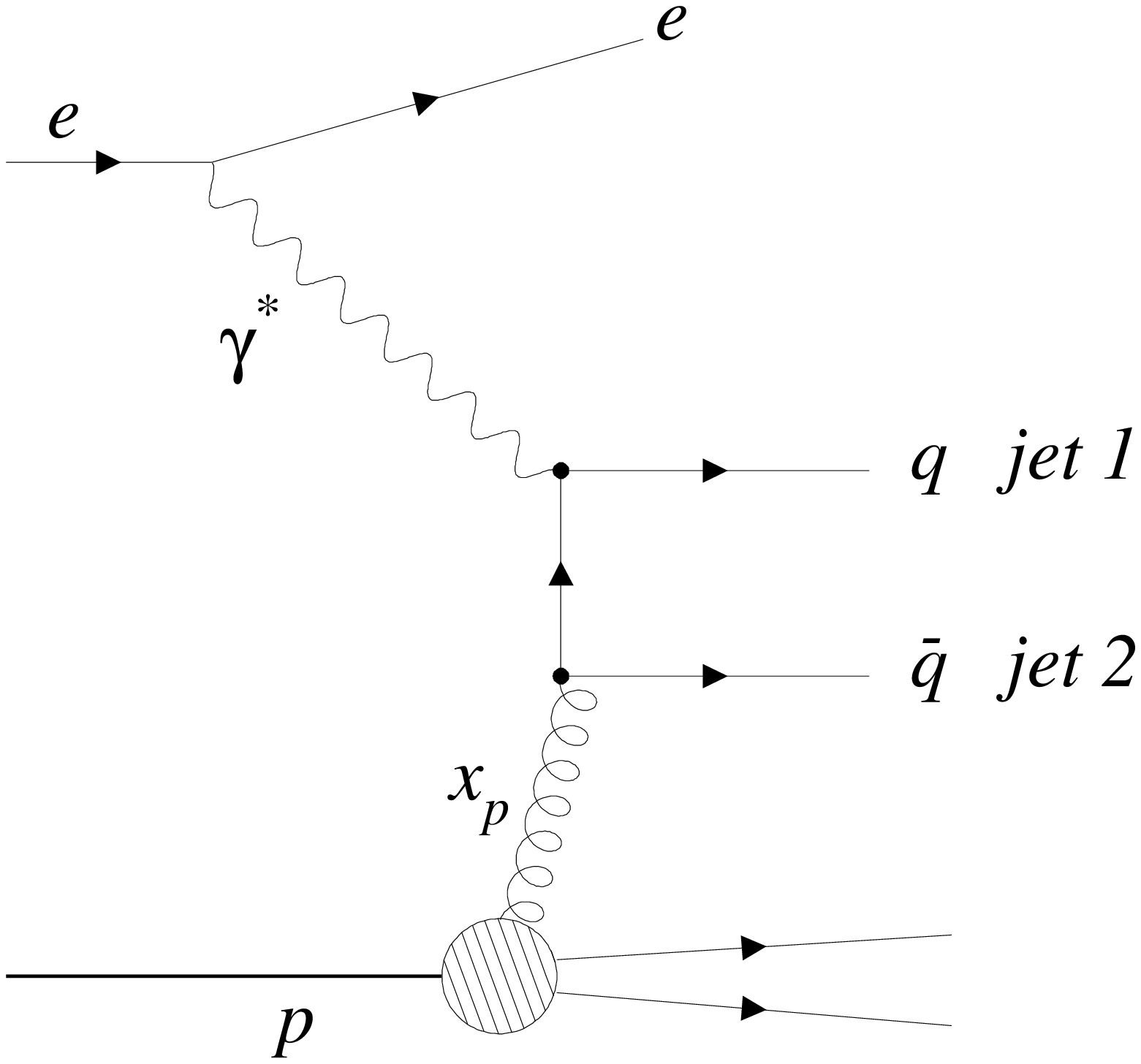,width=0.5\textwidth}\epsfig{file=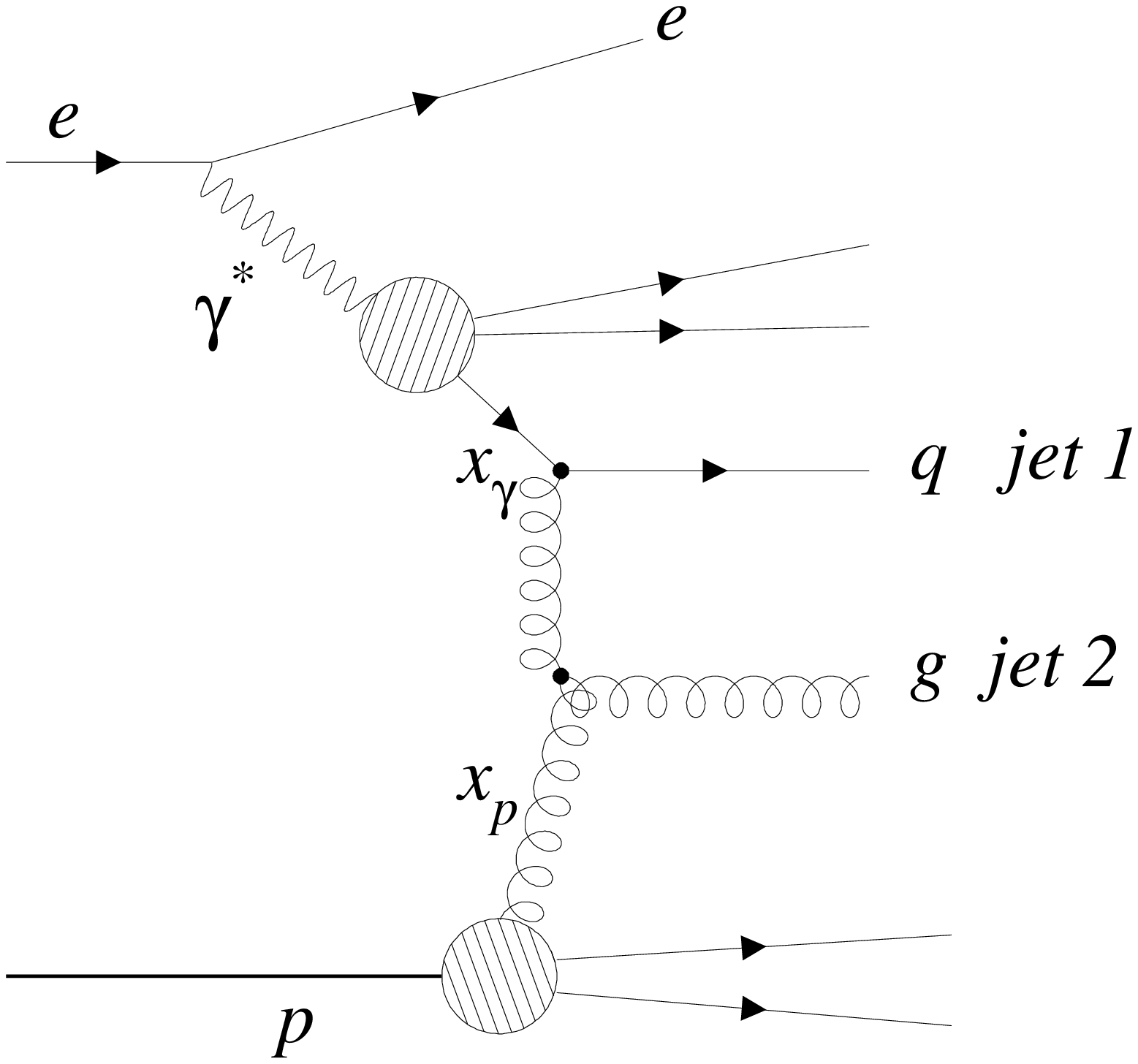,width=0.5\textwidth}
\end{center} 
\vspace*{-8cm}
a)\hspace{7.7cm}b)
\vspace{1.8cm}

\noindent
\hspace{4cm}$Q^2$

\noindent
\hspace{0.6cm}no structure $\uparrow$\hspace{6.2cm}structure $\uparrow$
\vspace{0.5cm}

\hspace{1.6cm}largest $P_T\updownarrow$\hspace{5.8cm}largest $P_T\updownarrow$
\vspace{1.9cm}

\hspace{1.3cm}structure $\downarrow$\hspace{6.2cm}structure $\downarrow$
\vspace{1.2cm}

\caption{\footnotesize Schematic diagram showing a) direct di--jet photoproduction in which the photon (real or virtual) couples directly to quark structure in the target proton, and b) resolved di--jet photoproduction in which the hadronic structure of the photon (real or virtual) is probed by a parton (quark or gluon) from the proton in a high$-P_T$ partonic subprocess; in a) the probe momentum transfer scale $P_T^2$ is necessarily carried by the exchange of an off--shell quark and all the momentum of the photon participates in the parton process; in b) the probe momentum transfer scale $P_T^2$ is carried by the exchange of an off--shell quark, or an off--shell gluon, and only a fraction $x_{\gamma}$ of the photon participates in the parton process.}
   \label{fig:dijets}
\end{figure}

Following the principles set out in section~\ref{sec:manifestation}, it should be possible to resolve and observe photon structure in high energy photoproduction and electroproduction processes in which a ``hard'' parton interaction takes place and jets due to parton hadronisation are produced. The principles are outlined in figure~\ref{fig:dijets}. By observing the production of jets from the hard parton interaction in which the transverse momentum squared $P_T^2$ is much greater than any other scale in the process, including $Q^2$, then the effects of structure in {\em both} the proton {\em and} the photon should be manifest at the probe scale $P_T^2$. If $P^2_T$ can be taken as the factorisation scale (section~\ref{sec:manifestation}), then, given adequate a priori knowledge of proton structure and the validity of factorisation, photon structure (both real and virtual) can be probed with the partons, both quarks and gluons, in the proton.

The photoproduction of jets through a hard parton process differs from the production in hadron--hadron collisions~\cite{Mont} in one respect -- photon structure is fluctuation--driven and therefore the photon interacts either by means of its direct coupling to partons in the proton, or by means of partons in its structure which are resolved in the interaction. The terminology of ``direct'' and ``resolved'' processes is now established, though theoretically the distinction is not clear--cut. For, as explained in section~\ref{sub:pointlikestructure}, fluctuation--driven structure is inherently a part of the manifestation of a quantum at some resolution scale so that higher order effects, which may not be resolvable but which are of course nevertheless always there, make the distinction ambiguous~\cite{Chyla}.

\subsubsection{Measurement and Kinematics}
Measurement of di--jet photoproduction requires the highest possible photon--proton interaction energy. This is available in electron/positron--proton ($ep$) collisions when $Q^2$ is much less than the mass squared $M^2_Z$ for which the effects of electroweak boson ($Z^0$) exchange are significant. Then it is possible to interpret the electroproduction of di--jets ($ep\rightarrow {\rm jet}+{\rm jet}+X$, $X$ being remnant particle production) in terms of photoproduction of di--jets for which the photon has space--like $Q^2$ over a wide range. If the factorisation scale $P_T^2$ exceeds $Q^2$, the di--jet production cross section will be sensitive to photon structure of spatial extent characterised by $Q^2$.

The highest energy $ep$ interactions are at the HERA collider at DESY, Hamburg. Di--jet production is measured by looking for interactions in which two jets of hadrons are produced, each observed as a cluster of energy in a ``calorimeter''. The jets are required to be produced with large ``transverse energy'' $E_T$ relative to the photon--proton direction in their centre--of--mass. The variable $E_T$ specifies the transverse momentum $P_T$ of the jets, and is so called because it specifies energy flow transverse to the photon--proton interaction, and because it is a measure of $P_T$.

The kinematics is straightforward, though a little more complicated than for DIS.

The photon is of course specified exactly as for $ep$ DIS by its virtuality $Q^2$ and its fractional momentum (inelasticity) variable $y$ for the incident electron splitting (following the splitting equations~\ref{eq:tdef} and~\ref{eq:y} in section~\ref{sec:kin})\footnote{\label{foot:PQconfusion}What has become a conventional choice of notation can cause confusion here when comparing the role of photon structure in high $E_T$ photoproduction with the same in deep--inelastic $e\gamma^*$ scattering, which is discussed in the first part of this paper. In photoproduction the structure of the virtual photon is probed by means of the partons in the proton, so that everywhere $Q^2$ specifies the size of the probed photon at factorisation scale $E^2_T$. In deep--inelastic $e\gamma^*$ scattering an electron probes a photon with mass squared $-P^2$ with a harder virtual photon at a factorisation scale $Q^2$.}. 

The factorisation scale is taken to be the jet $E_T$. Two inelasticities $x_{\gamma}$ and $x_{p}$, which are related to the rapidities $\eta_1$ and $\eta_2$ of each jet (relative to the photon direction) and to $E_T$ according to (figure~\ref{fig:dijets})
$$x_{\gamma}=\frac{p_1.P}{q.P}=\frac{E_Te^{\eta_1}+E_Te^{\eta_2}}{E_{\gamma}}$$
and
$$x_{p}=\frac{p_2.P}{q.P}=\frac{E_Te^{-\eta_1}+E_Te^{-\eta_2}}{E_p}$$
specify the incoming parton momenta. Here $p_1$ and $p_2$ are $4-$momenta of the partons from the photon and proton structure respectively, and the logarithmic relationship equation~\ref{eq:ytorapidity} clearly is the basis of this expression. There are more technical issues concerning the possibility of extra jets and interactions between partons in the remnants of the interacting photon and hadron which complicate the analysis~\cite{H1photojets}.

\subsubsection{Differential Cross Section for Di--jet Production}
\begin{figure}[hbt]
\begin{center}
\epsfig{file= 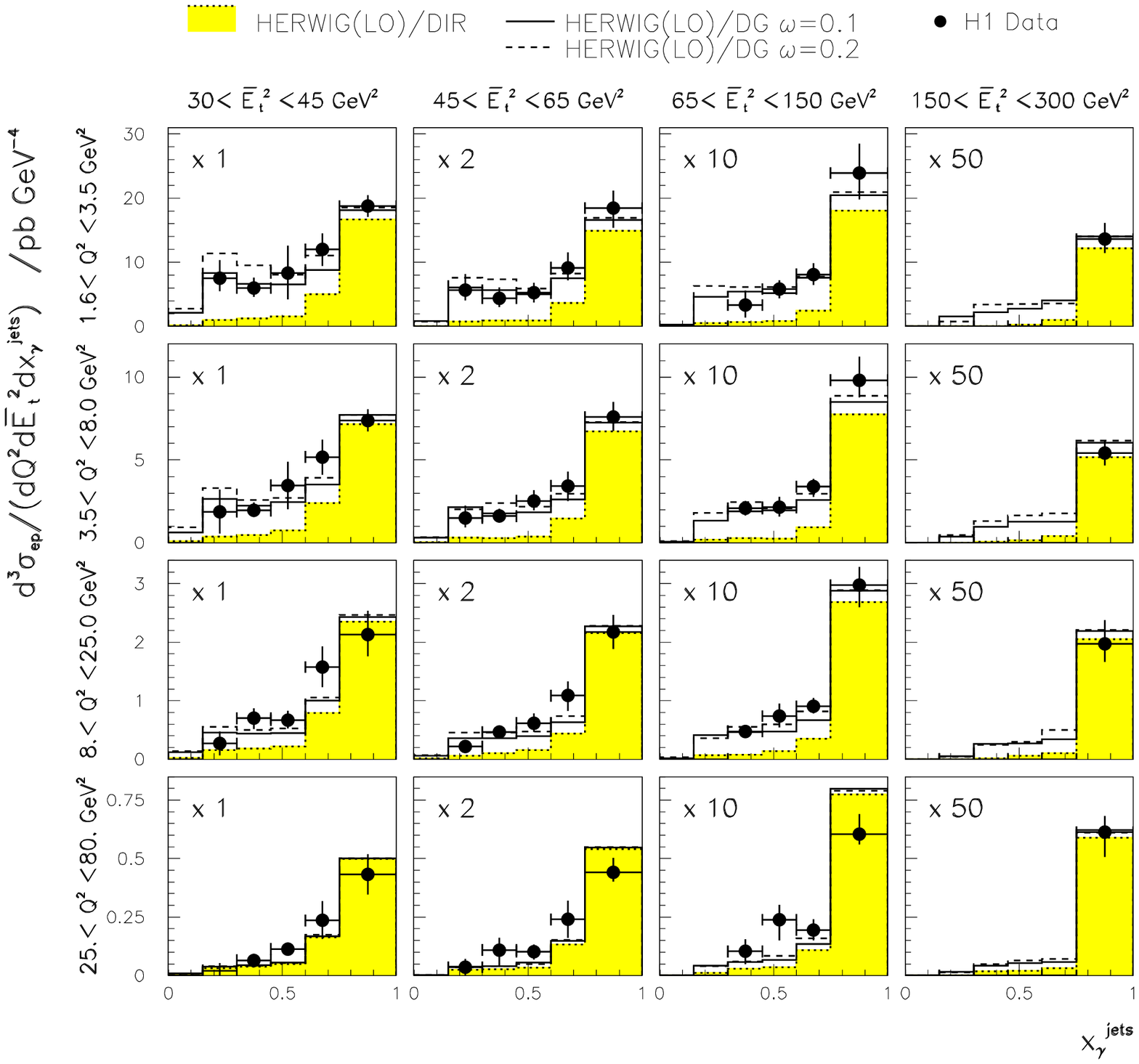,width=0.7\textwidth}
\end{center}
\caption{\footnotesize Cross section ${\rm d}^3\sigma_{ep}/{\rm d}Q^2{\rm d}E^2_T{\rm d}x_{\gamma}$ as a function of $x_{\gamma}$ for different virtual photon mass squared $Q^2$ and different probe scale $E_T^2$; direct interactions have $x_{\gamma}\rightarrow 1$; virtual photon structure is visible provided the probe scale $E_T^2$ is sufficient to resolve structure within the photon of size $\propto 1/Q$; the predictions HERWIG(LO)/DIR, HERWIG(LO)/DG($\omega=0.1$), HERWIG(LO)/DG$(\omega=0.2$)~\cite{HERWIG}, correspond to a specific prediction based on a model for virtual photon structure; the measurement is by the H1 experiment at HERA~\cite{H1virtphotojets}.}
   \label{fig:gammavirtx}
\vspace*{-9cm}

\noindent
\hspace{15cm}$\uparrow$

\noindent
\hspace{14.4cm}target

\noindent
\hspace{14.4cm}bigger
\vspace{1cm}

\noindent
\hspace{14.9cm}$Q^2$

\noindent
\hspace{15cm}$\downarrow$
\vspace{2.5cm}

\hspace{6.5cm}probe resolution improving: $E^2_T$ $\rightarrow$
\vspace{2.6cm}
\end{figure}

The most striking feature of the cross section for di--jet production is the measured dependence on $x_{\gamma}$ (figure~\ref{fig:gammavirtx})~\cite{H1virtphotojets}. As expected there is a distribution reflecting a mix of direct and resolved processes for which there is no perfect separation experimentally. The data shown are taken from virtual photon di--jet production for which the range of $Q^2$ is quite large. The dependence on $x_{\gamma}$ is shown for different photon virtuality, that is for different photon size. Direct interactions are manifest in the region $x_{\gamma}\rightarrow 1$, resolved interactions elsewhere. ``Squeezing'' the photon by going to higher $Q^2$ shows how resolved interactions are only significant when as expected $E^2_T>Q^2$ and the spatial extent of the photon exceeds the resolution of the probe, just as for $e\gamma^*$ DIS measurements but now over a much larger range of photon size (section~\ref{sec:virtualphotonstructure}).

\begin{figure}[hbt]
\begin{center}
\epsfig{file=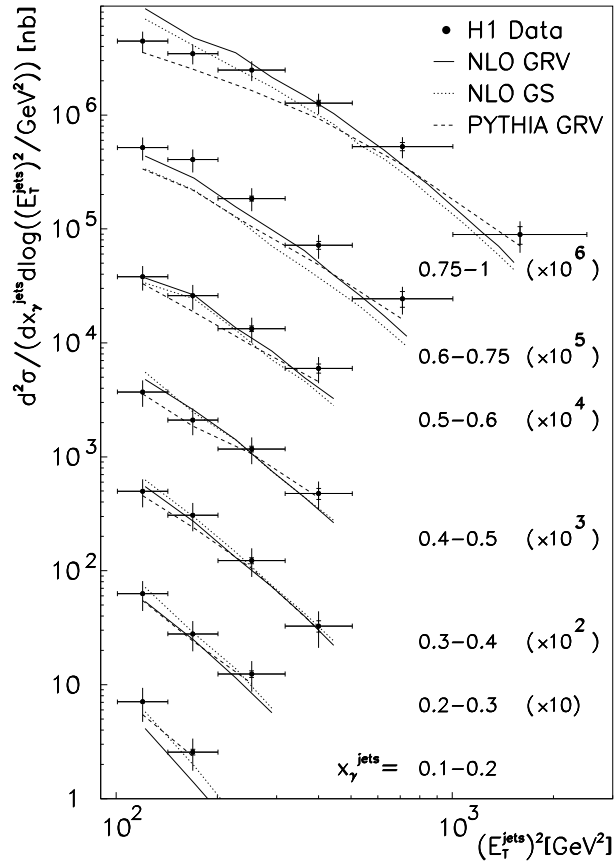,width=0.5\textwidth}
\end{center}
\caption{\footnotesize Cross section ${\rm d}^2\sigma_{ep}/{\rm d}E^2_T{\rm d}x_{\gamma}$ for di--jet photoproduction for nearly real photons ($Q^2\sim 0$) as a function of  $E_T^2$ for different $x_{\gamma}$; direct interactions have $x_{\gamma}\rightarrow 1$ and require an off-shell quark; resolved interactions are predominantly attributable to the exchange of off--shell gluons; the curves shown superimposed use the PYTHIA Monte Carlo simulation~\cite{PYTHIA} with GRV LO~\cite{GRVLO} pdfs for the photon (dashed), and analytical NLO calculations of di--jet cross sections~\cite{Klasen} using GRV HO QCD pdfs (continuous)~\cite{GRVHO} and GS96 HO QCD pdfs (dotted)~\cite{GS96}.}
   \label{fig:dijetxsection}
\end{figure}

The di--jet production differential cross section also tends to reveal very beautifully the QCD dynamics at work in the process. Figure~\ref{fig:dijetxsection} shows the $E_T^2$ dependence for different $x_{\gamma}$ for very nearly real photoproduction ($Q^2\sim 0$)~\cite{H1photojets}. An evolution is visible from a less to a more steep dependence as one moves away from the region $x_{\gamma}=1$ where direct interactions dominate. In direct interactions the full momentum of the photon participates in the hard parton interaction so that $E^2_T$ resides in the exchanged off-shell quark to which the photon must couple, as in figure~\ref{fig:dijets}a). The parton cross section $\frac{{\rm d}\hat{\sigma}}{{\rm d}\hat{t}}$  then follows a $\frac{1}{|\hat{t}|\hat{s}}$ dependence ($\hat{s}$ and $\hat{t}$ are the centre of mass energy squared and $4-$momentum transfer squared of the hard parton interaction in QCD: in LO $\hat{s}\sim x_{\gamma}x_p W^2$ and $\hat{t}\sim E^2_T$) which is expected with an off--shell quark propagator, so that one observes $\sim 1/E^2_T$ dependence in the jet cross section. For resolved interactions at smaller $x_{\gamma}$, $E^2_T$ can reside in either an off--shell quark or an off--shell gluon. Parton cross sections with an exchanged gluon follow a $\frac{{\rm d}\hat{\sigma}}{{\rm d}\hat{t}}\sim \frac{1}{|\hat{t}|^2}\sim \frac{1}{E_T^4}$ dependence (like Rutherford scattering where the virtuality is carried instead by an off--shell photon) in which there is no $\hat{s}$ dependence. Therefore at high $W^2$, and thereby high $\hat{s}$, the gluon exchange diagrams prevail, so that resolved processes follow a steeper $\sim 1/E^4_T$ dependence. Acknowledging also the influence of a decreasing kinematic reach in $E_T$ with decreasing $x_{\gamma}$, this tendency is visible in figure~\ref{fig:dijetxsection} at lower $x_{\gamma}$.

\subsubsection{Extracting Photon Structure}
The dominance of gluon exchange diagrams at high $\hat{s}$ in resolved processes in jet photoproduction can be exploited to extract the contribution of photon structure to the process. Combridge and Maxwell~\cite{CombridgeMaxwell} pointed out that the ubiquity of gluon exchange at large $\hat{s}$ means that the cross section for high$-P_T$ jet production could in practice be well approximated in a single expression, the single effective sub--process (SES). The procedure was first used in the 1980s in high energy $\bar{p}p$ physics to extract a direct measurement of the gluon density in the proton~\cite{UA1gluon}. 

Taking a LO view in QCD of di--jet photoproduction, the $ep$ cross section is written
\begin{equation}
\frac{{\rm d}^5\sigma}{{\rm d}y{\rm d}x_{\gamma} 
                       {\rm d}x_p {\rm d}\cos\theta^*  {\rm
                       d}Q^2} =   
  \frac{1}{32 \pi s_{ep}} 
\sum_{k={\rm T},{\rm L}}\frac{f_{\gamma/{\rm e}}^k(y,Q^2)}{y} 
\sum_{ij} 
\frac{f_{i/\gamma}^k(x_\gamma,P_T^2,Q^2)}{x_\gamma} 
\frac{f_{j/p}(x_{\rm p},P_T^2)}{x_p} 
|M_{ij}(\cos\theta^*)|^2
\label{eq:fullsig}
\end{equation}
where $f_{i/\gamma}^{\rm T}$, $f_{i/\gamma}^{\rm L}$ and $f_{i/p}$ are the pdfs (splitting functions for the photon, and parton densities for the proton) for each parton species $i$ in a transverse photon, a longitudinal photon 
and a proton respectively. They are evaluated at the factorisation 
scale
$P_T^2$. The $M_{ij}$ are QCD matrix elements for 
$2\rightarrow 2$ parton-parton
hard scattering processes. The quantity $s_{ep}$ is the square of the centre of mass
energy in the $ep$ collision, and $\theta^*$ is the polar angle of the outgoing
partons in the parton-parton centre of mass frame.
The fluxes of
transverse (${\rm T}$)
and longitudinal (${\rm L}$) photons are given by equations~\ref{eq:photonfluxes} but now of course with $Q^2$ replacing $P^2$ as the virtuality (footnote~\ref{foot:PQconfusion})~\cite{Budnev}.

In equation~\ref{eq:fullsig}, it is not possible from a measurement of just the dijet cross-section to disentangle the parton pdfs by separating 
the various parton sub-processes and the sensitivity to photon 
polarisation. To obtain a factorisable expression, a set
of polarisation-averaged parton densities of the photon are defined as follows:
$$f_{i/\gamma}(x_\gamma,P_T^2,Q^2) \equiv 
f_{i/\gamma}^{\rm T}(x_\gamma,P_T^2,Q^2)+\overline{\varepsilon}f_{i/\gamma}^{\rm L}(x_\gamma,P_T^2,Q^2),$$
where $\overline{\varepsilon} \sim 1$ is the ratio of longitudinal to
transverse photon fluxes, averaged over the $y-$range of the data. 
The pdfs $f_{i/\gamma}^{\rm L}$ are expected to be small over most of the kinematic 
range considered here (because of the predominance of helicity $\pm 1$ coupling in $\gamma\rightarrow q\bar{q}$ splitting)~\cite{vpth2,vpth6,frixetal}.
Then the SES approximation is used to replace the sum over parton processes by an SES cross-section
and effective parton densities for the photon and proton. Equation~\ref{eq:fullsig} can now be written to good approximation as 
\begin{eqnarray}
\frac{{\rm d}^5\sigma}{{\rm d}y  {\rm d}x_{\gamma} 
                       {\rm d}x_{\rm p}  {\rm d}\cos\theta^*  {\rm
                       d}Q^2} \approx 
 \frac{1}{32 \pi s_{ep}} 
\,\, \frac{f_{\gamma/{\rm e}}^T(y,Q^2)}{y}  
\,\, \frac{\tilde{f}_{\gamma}(x_\gamma,P_T^2,Q^2)}{x_\gamma} 
\,\, \frac{\tilde{f}_{{\rm p}}(x_{\rm p},P_T^2)}{x_{\rm p}}  
\,\, |M_{\rm SES}(\cos\theta^*)|^2\nonumber
\label{eq:SESsig}
\end{eqnarray}
where the effective parton densities are
\begin{eqnarray}
\tilde{f}_{\gamma}(x_\gamma, P_T^2,Q^2) & \equiv &
\sum_{\rm n_f} \left[f_{q/\gamma}(x_\gamma, P_T^2,Q^2)
               +f_{\overline{q}/\gamma}(x_\gamma, P_T^2,Q^2)\right] +
\frac{9}{4} \,  f_{g/\gamma}(x_\gamma, P_T^2,Q^2)\nonumber
\\
\tilde{f}_{\rm p}(x_{\rm p}, P_T^2) & \equiv &
\sum_{\rm n_f} \left[f_{q/p}(x_{\rm p}, P_T^2)
               +f_{\overline{q}/p}(x_p, P_T^2)\right] +
\frac{9}{4} \,  f_{g/p}(x_p, P_T^2)
\label{eq:effgamma} 
\end{eqnarray}
and the summations are over the quark flavours. Note the explicit occurrence of the colour factor $\frac{9}{4}$ to account for gluon scattering as part of the hard parton interaction.

\subsubsection{Photon Structure in Di--Jet Production}
\begin{figure}[hbt]
\begin{center}
\epsfig{file=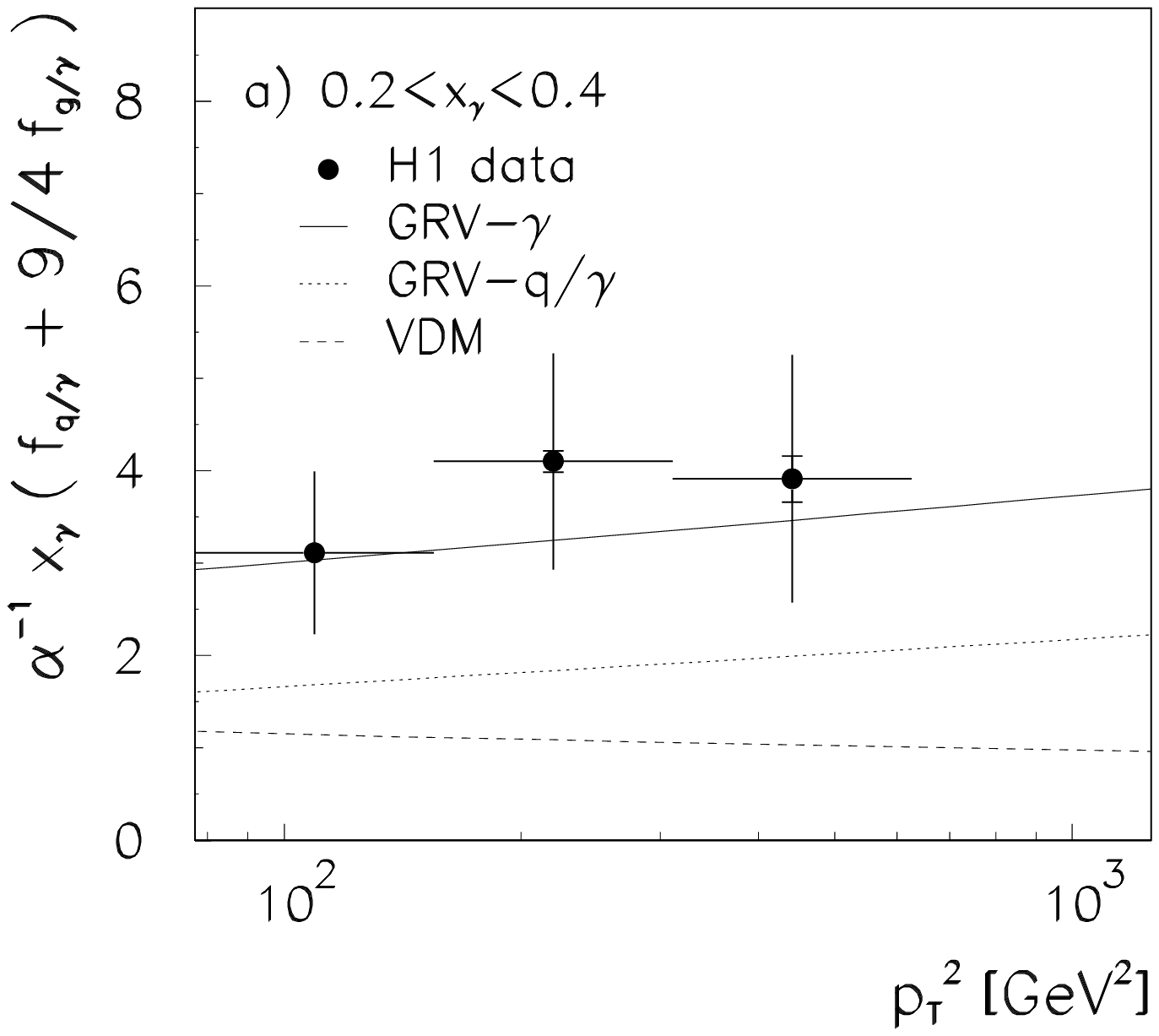,width=0.5\textwidth}\hfill\epsfig{file=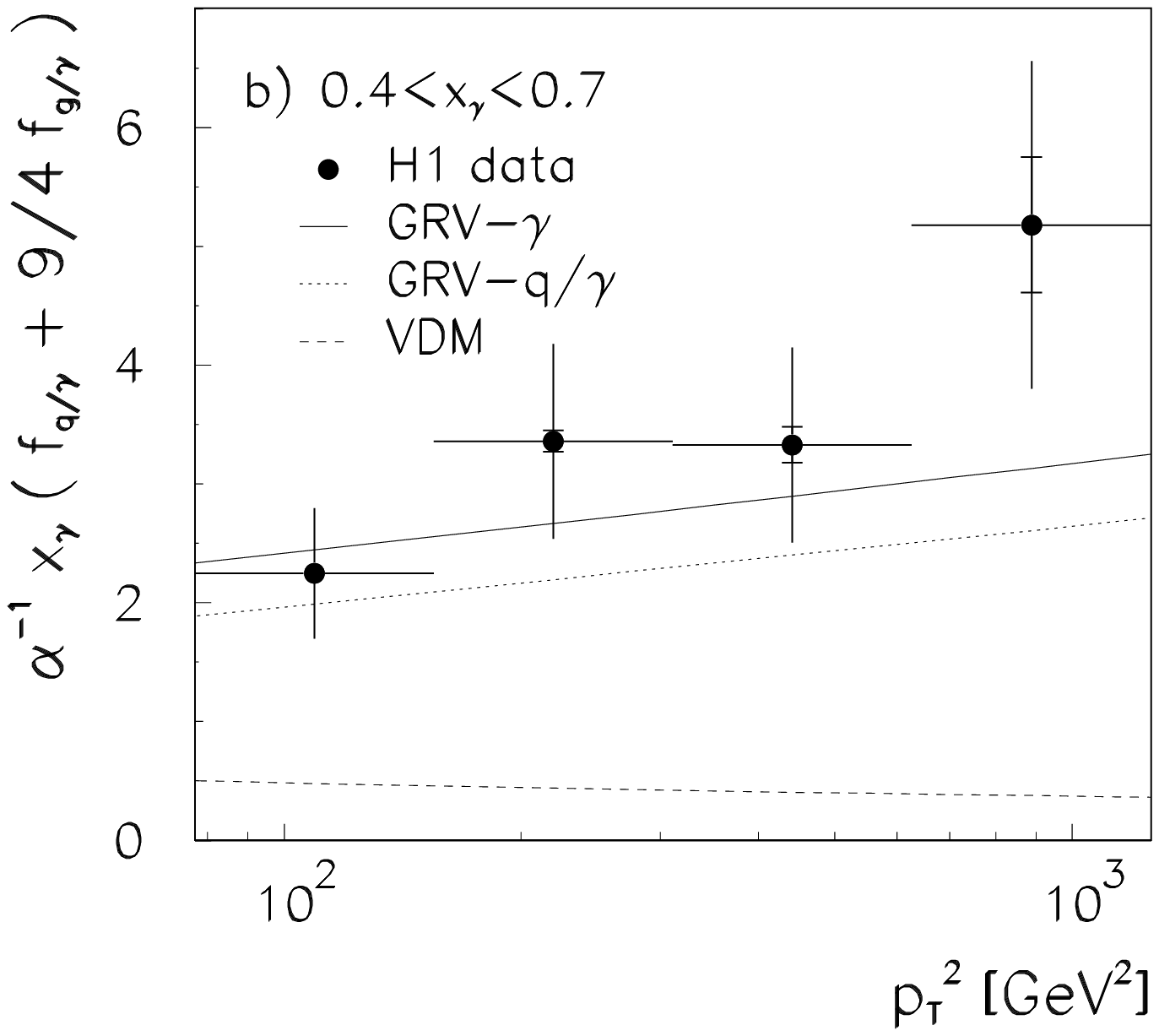,width=0.5\textwidth}
\end{center}
\caption{\footnotesize A measurement of the $P_T^2$ (or equivalently $E_T^2$) dependence of the effective parton density $F^{\gamma}_{\rm epdf}$ of the real photon extracted from di--jet photoproduction for two ranges of $x_\gamma$, a) $0.2<x_\gamma<0.4$ and b) $0.4<x_\gamma<0.7$; the continuous curve is the expectation assuming QCD LO photon pdfs~\cite{GRVLO}; the dotted curve shows the contribution to this GRV expectation from just the quarks in the photon, and the dashed line shows the contribution from VDM in this model; the measurement is by the H1 experiment at HERA~\cite{H1photojets}.}
   \label{fig:effgammaQ2}
\end{figure}

Once extracted from the measured cross sections, with the above assumptions and in the SES approximation, the effective parton densities (equation~\ref{eq:effgamma}) constitute an estimate of the pdfs of the photon evaluated in LO QCD. Figure~\ref{fig:effgammaQ2} shows a measurement of the $E_T^2$ dependence of the combination
$$F^{\gamma}_{\rm epdf}=\frac{1}{\alpha}x_\gamma\bigl(\sum_{\rm n_f} f_{q/\gamma}+\frac{9}{4}f_{g/p}\bigr)$$
which quantifies the structure in LO QCD associated with the real photon, for two different ranges of $x_\gamma$~\cite{H1photojets}. It reveals a persistent rise with increasing factorisation scale for both $x_\gamma$ ranges which is consistent with that seen in $F_2^{\gamma}$ measurements (figure~\ref{fig:scaling}), and thereby it demonstrates beautifully the expected photon structure, driven by quantum fluctuation, playing its role in di--jet photoproduction.

\begin{figure}[hbt]
\begin{center}
\epsfig{file=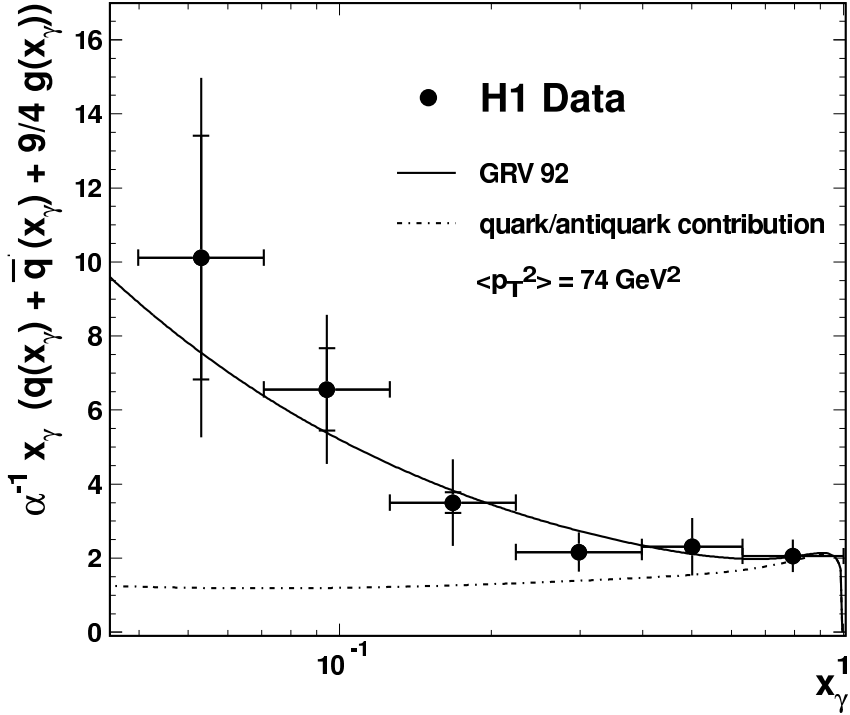,width=0.7\textwidth}
\end{center}
\caption{\footnotesize A measurement by the H1 experiment at HERA~\cite{H1photojets} of the $x_{\gamma}$ dependence of the effective parton density $F^{\gamma}_{\rm epdf}$ of the real photon extracted from di--jet photoproduction; the curves superimposed are the expectation from the QCD LO GRV92 parametrisation of photon structure~\cite{GRVLO}.}
   \label{fig:effgammax}
\end{figure}

Unlike measurements of $F_2^{\gamma}$ in $e\gamma$ DIS where the probe only couples to quarks, $F^{\gamma}_{\rm epdf}$ is directly sensitive to the gluon density in the photon. Figure~\ref{fig:effgammax} shows a measurement of the $x_\gamma$ dependence at factorisation scale $74\,{\rm GeV}^2$ which spans a large range of $x_\gamma$~\cite{H1lowxFeff}. The rather featureless $x-$dependence above $x_\gamma=0.2$, which is characteristic of the $e\gamma$ DIS measurements of $F_2^{\gamma}$ and which is driven by $\gamma\rightarrow q\bar{q}$ splitting, is observed. However at low $x_\gamma$ there is a sharp rise of $F^{\gamma}_{\rm epdf}$ because of the contribution of gluons. There is thus evidence that photon structure carries with it a gluon content which grows in a manner similar to that observed in proton structure at low$-x$~\cite{Foster}. This is expected in pQCD, as is clear from the contributions to $F^{\gamma}_2$ in figure~\ref{fig:QCDdiags} involving gluons at higher order, and from the incalculable ``hadron'' contribution, which, if it is indeed VMD, will have a growing gluon density much like that of the proton with increasing factorisation scale~\cite{Foster}.

\begin{figure}[hbt]
\begin{center}
\epsfig{file=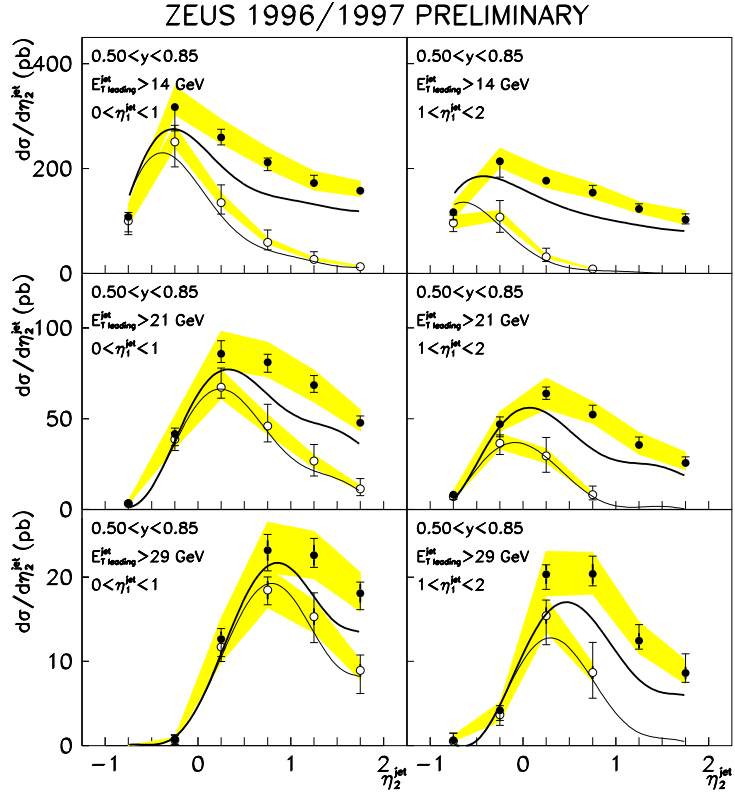,width=0.6\textwidth}
\end{center}
\caption{\footnotesize A measurement of the cross section ${\rm d}\sigma/{\rm d}\eta$ for di--jet production as a function of $\eta$ of the more energetic jet for different ranges of transverse energy $E_T$; the measurement is by the ZEUS experiment at HERA~\cite{ZEUSphotojets}; the filled data points are selected as predominantly resolved interactions and the open data points as direct interactions following an $x_{\gamma}$ based selection; shaded area shows the systematic uncertainty due to calorimeter energy scale; the thin(thick) curves show the expectations for direct(resolved) contributions in a NLO calculation following the same $x_{\gamma}$ based selection.}
   \label{fig:zeus_jet}
\end{figure}

What is also clear from the dependence on factorisation scale in figure~\ref{fig:effgammaQ2} is the reach of di--jet photoproduction measurements at HERA. This highlights how, when more data are taken by the HERA experiments, they will probe with precision a new domain of photon structure beyond $1000\,{\rm GeV}^2$ factorisation scale. Already there are strong hints from first measurements of the cross section for di--jet production at the highest $E_T$ that models built on analyses of photon structure from presently available data are not adequate. Figure~\ref{fig:zeus_jet} shows expectations compared with measured cross sections, where it can plainly be seen that the predictions undershoot the resolved photon contribution over much of the measured phase space~\cite{ZEUSphotojets}.

\begin{figure}[hbt]
\begin{center}
\epsfig{file=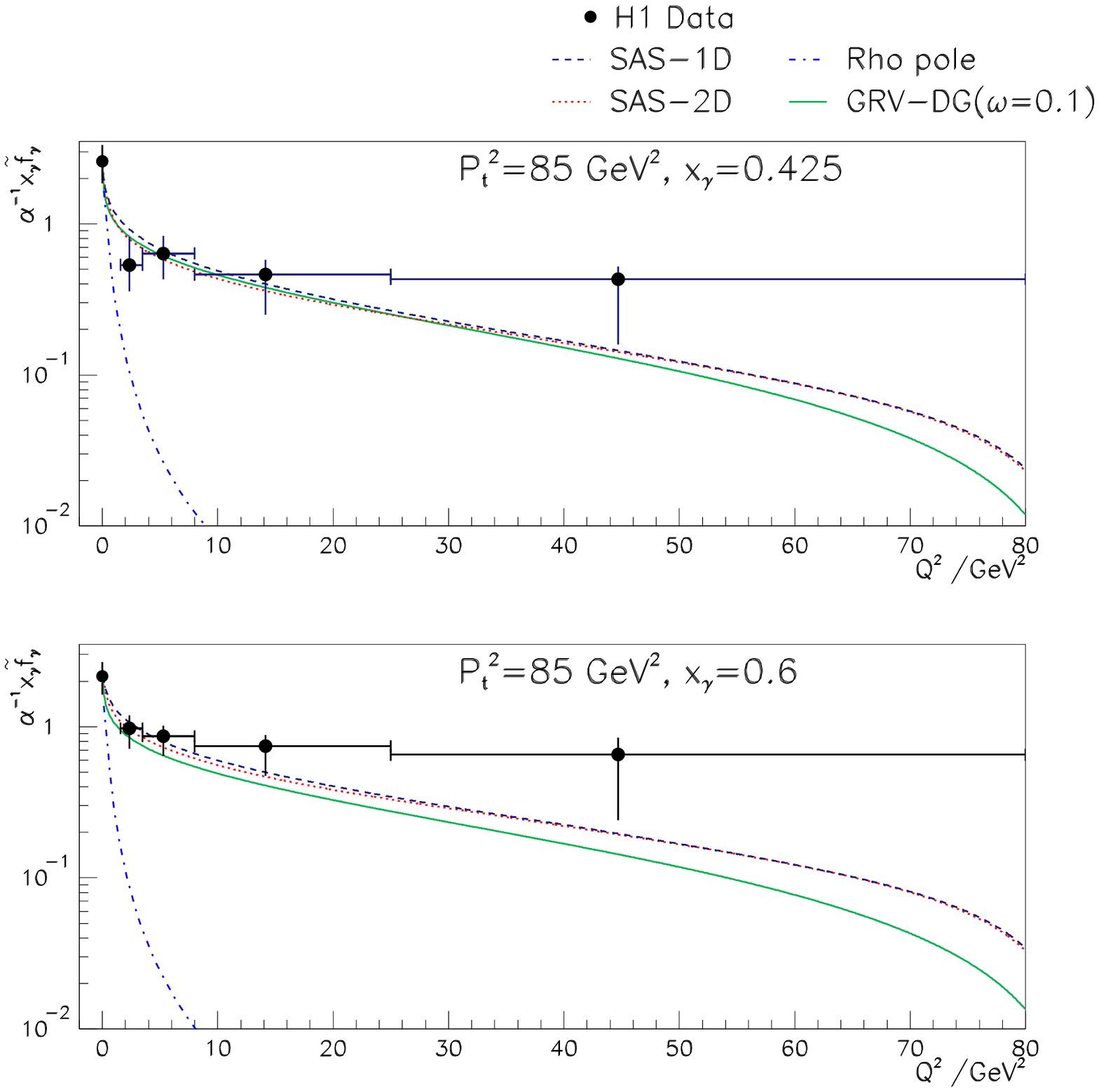,width=0.6\textwidth}
\end{center}
\caption{\footnotesize The $Q^2$ (target mass squared) dependence of the effective parton density $F^{\gamma}_{\rm epdf}$ (see text) measured in the electroproduction of di--jets for different $x_{\gamma}$ at a factorisation scale of $P_T^2=85\,{\rm GeV}^2$; theoretical curves are shown as follows: SAS-1D, SAS-2D~\cite{SAS} and GRV-DG($\omega=0.1$)~\cite{DreesGodbole} are parametrisations of photon structure based on QCD analyses of DIS $e\gamma$ measurements of nearly real (mass--less) photons; Rho pole shows the contribution expected from VMD (or HAD) contributions if dominated just by the $\rho(770)$. The decrease of $F^{\gamma}_{\rm epdf}$ by a factor of 2 or 3 at low $Q^2$ is clearly attributable to the demise with photon size of VMD, and the persistence of $F^{\gamma}_{\rm epdf}$ out to large $Q^2$, that is small size, to pQCD contributions.}
   \label{fig:H1_effpdf_virtual_photon_P2}
\end{figure}

The $Q^2$ dependence of $F^{\gamma}_{\rm epdf}$ is shown in figure~\ref{fig:H1_effpdf_virtual_photon_P2}~\cite{H1virtphotojets}. The substantial increase in kinematic reach of di--jet photoproduction measurements compared with DIS measurements of virtual photon structure (figure~\ref{fig:P2dependences}) shows clearly how, as already noted (section~\ref{sec:virtualphotonstructure}), the VMD contribution (in its commonly accepted theoretical estimate -- Rho pole) quickly dies away with decreasing photon size so that beyond $Q^2\sim 3\,{\rm GeV}^2$ only the perturbatively calculable contribution in equation~\ref{eq:LOQCD} survives. 
This is a remarkable result. For the first time one sees the evolution of hadronic structure from a region ($Q^2=0\,{\rm GeV}^2$) where through VMD the size of the hadronic system is that which is typical of hadrons, namely $1$ fm, and where pQCD inevitably fails, to a region ($Q^2\sim 45\,{\rm GeV}^2$) where the hadronic system has dimension $\propto \frac{1}{Q}$ (equation~\ref{eq:transversesize}) of about $0.03$ fm and where pQCD can provide a complete prediction. Understanding this evolution is crucial to understanding in QCD the transition between hadronic physics which can be calculated perturbatively -- jet/parton dynamics, and hadronic physics which requires a non-perturbative approach despite the fact that the degrees of freedom are still those of asymptotically free partons -- hadron valence structure. De facto this in turn will quantify the evolution of the colour dipole which in some approachs can form a basis of an understanding of low$-x$ electron--proton DIS, and thereby will help to elucidate the distinction between effects due to proton structure and effects due to the transition in QCD between soft and hard hadronic physics~\cite{Foster}.

Furthermore, though the present precision of the measurements is limited (figure~\ref{fig:H1_effpdf_virtual_photon_P2}), it is already clear that QCD predictions of the structure of a virtual photon with space--like mass squared above about $30\,{\rm GeV}^2$ may well undershoot the measurements. Better precision is required, and will come. 

\subsubsection{Summary}

Measurements of jet electroproduction at HERA, and their interpretation in terms of real and virtual photoproduction of jets, confirm the simple picture of photon structure first identified in $e\gamma$ DIS, and make possible measurements of this structure over a substantially wider kinematic range. In so doing they demonstrate the validity of a factorisable structure which can be assigned to a photon, both  real and virtual. The kinematic reach of these HERA measurements far outstrips that of present $e^+e^-$ experiments at LEP.

\section{Conclusion}

Photon structure has always been central, either implicitly or explicitly, to major steps in our understanding of the physics of short distances and times. Its influence in Particle Physics stems from the discovery of the positron, the development of QED, and the impact of quantum fluctuation on the behaviour of otherwise point--like quanta. Latterly this has taken the form of a means of testing QCD by probing the spatially extended structure of the photon ($\gamma$) due to hadronic fluctuation. 

A simple picture emerges of a factorisable hadronic structure of the photon which is driven by quark--antiquark ($q\bar{q}$) pair production and chromodynamics. The characteristic features of the structure function of the photon $F_2^{\gamma}$ are a logarithmic growth with increasing factorisation scale and a relatively featureless and slightly rising dependence on Bjorken$-x$. The former is attributable to the splitting function ${\cal P}_{q/\gamma}$ in QCD. The latter can be understood in terms of the competing influences of the rising $x-$dependence in perturbative QCD of ${\cal P}_{q/\gamma}$ and the falling $x-$dependence of a VMD contribution, which is to date incalculable in QCD and which is interpreted as being due to soft hadronic fluctuation of the photon to vector mesons. 

However this picture continues to pose a serious challenge to QCD because it relies on phenomenology and is thereby quantitatively incomplete. This challenge may well be met with the symbiosis of QCD theory and improving measurements of the structure of increasingly compact photons. Here the presence of two hard scales, that of the probe and that of the target, facilitates QCD calculation. This in turn may lead to new phenomena and insight as the short--distance properties of the QCD fluctuation of virtual photons are forced to ever smaller dimension and are probed with ever greater precision.

\section*{Acknowledgements}

All the results presented here are due to the work of many in the collaborations that form the basis of modern High Energy Physics. I wish to thank numerous colleagues who over the years have influenced and helped my doing, thinking and understanding while working on the PLUTO experiment at the PETRA $e^+e^-$ storage rings, and on the H1 experiment at the HERA $ep$ collider. I wish also to thank the Royal Society for its hospitality, and I Butterworth, J Ellis, and E Gabathuler for their organisation of an especially stimulating meeting. My involvement in the curiosities of lepton/photon physics and photon structure has been maintained with the support for many years of the UK Science Research Council, the UK Science and Engineering Research Council, the UK Particle Physics and Astronomy Research Council, the Universities of Glasgow and Liverpool, and the DESY Laboratory, Hamburg, Germany, for which I am grateful.

\end{document}